\documentclass{llncs}
\usepackage{latexsym,color,graphics,times}
\usepackage{diagrams}
\usepackage{url}
\usepackage{epsfig}
\usepackage{amssymb}
\usepackage{amsmath}
\usepackage{amsfonts}

\begingroup
\catcode`\~=11
\gdef\urltilde{\lower 0.6ex\hbox{~}}
\endgroup

\newcommand{\A}{\mathcal{A}} \newcommand{\B}{\mathcal{B}}
\newcommand{\C}{\mathcal{C}} \newcommand{\D}{\mathcal{D}}
 
\newcommand{\G}{\mathcal{G}} 
\newcommand{\I}{\mathcal{I}} 
 \renewcommand{\L}{\mathcal{L}}
\newcommand{\M}{\mathcal{M}} 
 \renewcommand{\P}{\mathcal{P}}
 
\renewcommand{\S}{\mathcal{S}}

\newcommand{\tup}[1]{\langle #1\rangle}            

\newcommand{\ext}[2]{#1^{#2}}

\newcommand{\arity}[1]{\mathit{arity}(#1)}

\newcommand{\ins}[1]{\mathbf{#1}}

\newcommand{\key}{\mathit{key}}
\newcommand{\can}{\mathit{can}}
\newcommand{\ret}{\mathit{ret}}

\newcommand{\expand}[1]{\mathit{exp}_\G(#1)}

\newcommand{\unfold}[1]{\mathit{unf}_{\M}(#1)}

\newcommand{\Map}[2]{#1\,\leadsto\,#2}

\title{DB Category: Denotational Semantics for View-based Database Mappings}
\date{}

\author{Zoran Majki\'c}
\institute{ \email{majk.1234@yahoo.com}\\
~~~~http://zoranmajkic.webs.com/}

\newtheorem{theo}{Theorem}
\newtheorem{propo}{Proposition}
\begin{document}


\maketitle
\begin{abstract}
 We present a categorical denotational semantics for a
database mapping, \emph{based on views}, in the most general
framework of a database integration/exchange.
 Developed  database category $~DB~$, for databases (objects) and view-based mappings (morphisms) between
 them,
 is different from $Set$ category: the morphisms (based on a set of complex query computations) \emph{are not} functions,
  while the objects are database instances (sets of relations). The
 logic based schema mappings between databases, usually written in a
 highly expressive logical language (ex. LAV, GAV, GLAV mappings, or tuple
 generating dependency) may be functorially translated into this
 "computation" category $DB$. A new approach is adopted, based on the
 behavioral point of view for databases, and  behavioral equivalences for databases and their
 mappings are
 established. By introduction of view-based observations for databases, which are
 computations without side-effects, we define a
 fundamental (Universal algebra) monad with a power-view endofunctor $T$.
  The resulting 2-category $DB$ is symmetric,  so that  any mapping can be represented as an object (database instance) as well,
  where a higher-level mapping between mappings is a 2-cell morphism.
   Database category $DB$ has the following properties: it is equal to its dual, complete
   and cocomplete.
   Special attention is devoted to  practical examples: a query
   definition, a query rewriting in GAV Database-integration
   environment,
    and  the fixpoint solution of a canonical data-integration model.
\end{abstract}

%

\section{Introduction}

Most work in the data integration/exchange and P2P framework is
based on a logical point of view (particularly for the integrity
constraints, in order to define the right models for  certain
answers) in a 'local' mode (source-to-target database), where a
general 'global' problem of a \emph{composition} of complex partial
mappings that involves a number of databases has not been given the
correct attention. Today, this 'global' approach cannot be avoided
because of the necessity of  P2P open-ended networks of heterogenous
databases. The aim of this work is a \emph{definition of category
$DB$ for database mappings} more suitable than a $Set$ category: The
databases are more complex structures w.r.t. sets, and the mappings
between them are too  complex to be represented by a single
(complete) function. Why do we need  an enriched categorical
semantic domain such as this for databases?
We will try to give a simple answer to this question:\\
- $~~~~$This work is an attempt to give a correct  solution for a
general problem of complex database-mappings and for high level
algebra operators for databases (merging, matching, etc.),
preserving the traditional common practice logical language for
schema database
mapping definitions.\\
- $~~~~$The query-rewriting algorithms are not integral parts of a
database theory (used to define a database schema with integrity
constraints); they are \emph{programs} and we need an enriched
context that is able to formally express  these programs trough
mappings between databases as well. \\
- $~~~~$ Let us consider, for example, P2P systems or mappings in a
complex Datawarehouse: formally, we would like to make a synthetic
graphic representations of database mappings and queries  and to
develop a graphic tool for a meta-mapping description of complex
(and partial) mappings in various contexts, with a formal
mathematical background. \\
Only a few works considered this general problem
~\cite{MBDH02,AlBe02,DaBK98,MeRB03}. One of them, which uses a
category theory ~\cite{AlBe02}, is too  restrictive: their
institutions can be applied only for  \emph{inclusion} mappings
between databases.\\
There is a lot of work for sketch-based denotational semantics for
 databases \cite{LeSp90,RoWo92,DiCa95,JoRo00}. But all of them
use, as objects of a sketch category, the elements of an ER-scheme
of a database (relations, attributes, etc..) and not the
\emph{whole} database as a single object, which is what we need in a
framework of inter-databases mappings. It was shown in \cite{Disc97}
that if we want to progress to more expressive sketches w.r.t. the
original Ehresmann's sketches for diagrams with limits and
coproducts, by eliminating non-database objects as, for example,
cartesian products of attributes or powerset objects, we need
\emph{more expressive arrows} for sketch categories (diagram
predicates in \cite{Disc97} that are analog to the approach of
Makkai in \cite{Makk94}). Obviously, when we progress to a more
abstract vision where objects are the (whole) databases, following
the approach of Makkai, in this new basic category $DB$ for
databases, where objects are just the database instances (each
object is a set of relations that compose this database instance),
we will obtain much more complex arrows, as we will see. Such arrows
are not simple functions, as in the case of base $Set$ category, but
complex trees (operads) of view-based mappings. In this way, while
Ehresmann's approach prefers to deal with few a fixed diagram
properties (commutativity, (co)limitness), we enjoy the possibility
of setting full
relational-algebra signature of diagram properties.\\
This work is an attempt to give a correct solution for this problem
while preserving the traditional common practice logical language
for the
schema database mapping definitions. Different properties of this DB category are considered in a number of previously
published papers \cite{Majk09a,Majk09f,Majk08mm,Majk11mm,MaBh10} as well.\\
This paper follows the following plan: In Section 2 we present an
Abstract Object Type based on view-based observations. In Section 3
we develop a formal definition for a Database category $DB$, its
power-view endofunctor, and its duality property. In Section 4 we
formulate the two equivalence relations for databases (objects in
$DB$ category): a strong and a weak observation equivalences.
Finally, in Section 5 we present an application of this theory to
the data integration/exchange systems, with an example for a
query-rewriting in  data integration system,
 and we define a fixpoint operator for an infinite canonical solution in data
integration/exchange systems.
\subsection{Technical Preliminaries}
\begin{figure}
\begin{center}
 \includegraphics{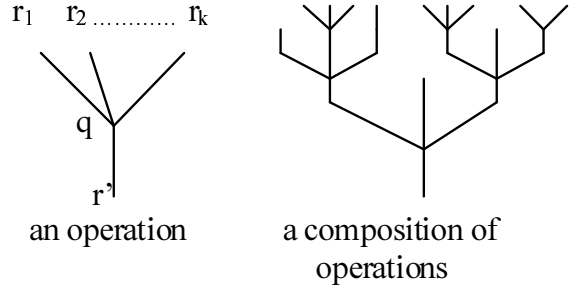}
 \caption{Operations of an R-operad}
 \label{Operations}
 \end{center}
 \end{figure}

The database mappings,  for a given logical language, are defined
usually at a schema level, as follows:
\begin{itemize}
  \item  A \emph{database schema} is a pair $\A = (S_h , S_n)$ where: $S_h$ is
  a countable set of relation symbols $r\in R$ with finite arity,
disjoint from a countable infinite set $\textbf{att}$ of attributes
(for any $\textbf{x} \in \textbf{att}$ a domain of $\textbf{x}$ is a
nonempty subset $dom(\textbf{x})$ of a countable set of individual
symbols $\textbf{dom}$, disjoint from $\textbf{att}$ ), such that
for any $r\in R$, the sort of $R$ is a finite sequence of elements
of $\textbf{att}$. $S_n$ denotes a set of closed formulas called
integrity constraints, of the sorted first-order language with sorts
$\textbf{att}$, constant symbols $\textbf{dom}$,
relational symbols $R$, and no function symbols.\\
A \emph{finite} database schema is composed by a finite set $S_h$,
so that the set of all attributes of such a database is finite.
\item An \emph{instance} of a database $\A$ is given by $A = (\A,I_A)$, where
$I_A$ is an interpretation function that maps each schema element of
$S_{h_A}$ (n-ary predicate) into an n-ary relation $a_i\in A $
(called also "element of  $A$" ). Thus, a relational
instance-database $A$ is a set of n-ary relations.
\item We consider a rule-based
\emph{conjunctive query} over a database schema $\A$ as an
expression $ q(\textbf{x})\longleftarrow R_1(u_1), ..., R_n(u_n)$,
where $n\geq 0$, $R_i$ are the relation names (at least one) in $\A$
or the built-in predicates (ex. $\leq, =,$ etc..), $q$ is a relation
name not in $\A$, $u_i$ are free tuples (i.e., may use either
variables or constants). Recall that if $v = (v_1,..,v_m)$ then
$R(v)$ is a shorthand for $R(v_1,..,v_m)$. Finally, each variable
occurring in $\textbf{x}$ must also occur at least once in
$u_1,...,u_n$. Rule-based conjunctive queries (called rules) are
composed by: a subexpression $R_1(u_1) , ...., R_n(u_n)$, that is
the \emph{body}, and $q(\textbf{x})$ that is the \emph{head} of this
rule. If one can find values for the variables of the rule, such
that the body holds (i.e. is logically satisfied), then one may
deduce the head-fact.
 This concept is captured by a notion of "valuation". In the
 rest of this paper a deduced head-fact will be called  "a resulting \emph{view}
 of a query $q(\textbf{x})$ defined over a database $\A$". Recall that the conjunctive
queries are monotonic and satisfiable. The $Yes/No$ conjunctive
queries are the rules with an empty head.
 \item We consider that a
\emph{mapping} between two databases $\A$ and $\B$ is expressed by
an union of "conjunctive queries with the same head". Such mappings
are called "view-based mappings".
 Consequently  we consider \emph{a
view} of an instance-database $A$ an n-ary relation (set of tuples)
obtained by a "select-project-join + union" (SPJRU) query
$q(\textbf{x})$  (it is a term of SPJRU algebra)  over
    $A$: if this query is a finite term of this algebra than it is
    called a "finitary view" (a finitary view can have also an infinite number of tuples).
\end{itemize}
We consider the views as a universal property for databases:
  they are the possible observations of the information contained in
  an  instance-database, and we may use them in order to establish an equivalence relation
  between databases.
      Database category $DB$, which will be introduced in what follows, is
      at an \emph{instance level}, i.e., any object in $DB$ is an instance-database
(i.e., a set of relations). The connection between a logical
(schema) level and this computational category is based on the
\emph{interpretation} functors. Thus, each rule-based conjunctive
query at schema level over a database $\A$ will be translated (by an
interpretation functor) in a morphism in $DB$,  from an
instance-database $A$ (a model of the database schema $\A$) to the
instance-database  composed by all views of  $A$.\\
    In what follows we will work with the typed operads, first developed for
    a purpose of homotopy theory ~\cite{Adam78,BoVo73,PMay68}, having a
set $R$ of types (each relation symbol is a type), or "R-operads"
for short. The basic idea of an R-operad $O$ is that, given types
$r_1,...,r_k, r\in R$, there is a set $O( r_1,..., r_k, r )$ of
abstract $k\_ary$ "operations" with inputs of type $r_1,...,r_k$ and
output of type $r$. We can visualize such an operation as a tree
with only one node. In an operad, we can obtain new operations from
old ones by composing them: it can be visualized in terms of trees
(Fig. \ref{Operations})
We can  obtain the new operators from old ones by permuting
arguments, and there is a unary "identity" operation of each type.
Finally, we insist on a few plausible axioms: the identity
operations act as identities for composition, permuting arguments is
compatible with composition, and \textit{composition is
associative}. Thus, formally, we have the following:
\begin{definition} \label{def:operads} For any set R, an R-operad O consists of
\begin{enumerate}
  \item for any $r_1,...,r_k,r \in R$, a set $O(r_1,..., r_k, r)$
  \item for any $f\in O(r_1,..., r_k, r)$ and any $g_1\in O(r_{11},...,r_{1i_1},
  r_1)$,..., $g_k\in O(r_{k1},...,r_{ki_k},r_k)$, an element
  $f\cdot(g_1,...,g_k)\in O(r_{11},..., r_{1i_1},...,r_{k1},..., r_{ki_k},r)$
  \item for any $r\in O$, an element $1_r\in O(r,r)$
  \item for any permutation $\sigma\in R_k$, a map
  $\sigma:O(r_1,..., r_k, r)\rightarrow O(r_{\sigma (1)},..., r_{\sigma (k)}, r)$,  $f\longmapsto f\sigma$, such that:
\begin{enumerate}
  \item whenever both sides make sense,
  $f\cdot (g_1\cdot(h_{11},...,h_{1i_1}),..,.g_k\cdot (h_{k1},...,h_{ki_k})) =
  (f\cdot (g_1,..,g_k))\cdot(h_{11},...,h_{1i_1},..,h_{k1},...,h_{ki_k})$
  \item for any $f\in O(r_1,..., r_k , r)$,     $~f = 1_r\cdot f =
  f\cdot(1_{r1},..., 1_{rk})$
  \item for any $ f\in O(r_1,..., r_k , r)$, and $ \sigma,
  \sigma_1\in R_k $, $~ f(\sigma \sigma_1) = (f\sigma)\sigma_1 $
  \item for any $ f\in O(r_1,..., r_k, r)$, $~\sigma\in R_k $
  and $g_1\in O(r_{11},..., r_{1i_1}, r_1)$,..., \\$g_k\in O(r_{k1},...,
  r_{ki_k},r_k)$, $~(f\sigma)\cdot((g_{\sigma (1)},..., g_{\sigma
  (k)}) = (f\cdot(g_1,..,g_k))\rho (\sigma) $  \\ where $\rho
  :R_k \longrightarrow R_{i_1 +...+i_k}$ is the obvious
  homomorphism.
  \item for any $ f\in O(r_1,..., r_k, r)$, $~g_1\in O(r_{11},..., r_{1i_1},
  r_1)$,..,$g_k\in O(r_{k1},..., r_{ki_k},r_k)$, and $ \sigma_1\in
  R_{i_1},....,\sigma_k\in R_{i_k}$, $~(f\cdot (g_1 \sigma_1,...g_k
  \sigma_k)) =
  (f\cdot(g_1,..,g_k))\varrho_1(\sigma_1,...,\sigma_k)$, \\ where
  $\varrho_1:R_{i_1}\times...\times R_{i_k} \longrightarrow R_{i_1
  +...+i_k}$ is the obvious homomorphism.
\end{enumerate}
\end{enumerate}
\end{definition}
Let us define the "R-algebra" of an operad where its abstract
operations are represented by actual functions (query-functions).
For a given database schema with relation symbols $r_1,..., r_k$ we
consider $ f\in O(r_1,..., r_k, r)$ as a conjunctive query $r
\leftarrow r_1,..., r_k$ that defines a view $r$.
\begin{definition}\label{def:operads2} For any R-operad O, a R-algebra $\alpha$
consists of:
\begin{enumerate}
  \item for any $ r\in R$, a set $ \alpha(r)$ is a set of tuples of
  this type (relation).
   $~\alpha^*~$ is the extension of $\alpha~$ to a list of symbols
  $~\alpha^*(\{r_1,..., r_k \}) \triangleq \{\alpha(r_1),...,
  \alpha(r_k)\}$.
  \item for any $q \in O(r_1,..., r_k, r)$ a mapping function
  $ \alpha(q):\alpha(r_1)\times...\times \alpha(r_k)
  \longrightarrow\alpha(r)$, such that
\begin{enumerate}
  \item whenever both sides make sense,
  $\alpha(q\cdot(q_1,..,q_k)) =
  \alpha(q)(\alpha(q_1)\times...\times\alpha(q_k))$
  \item for any $r\in R$, $\alpha(1_r)$ acts as an identity on
  $\alpha(r)$
  \item for any $q\in O(r_1,..., r_k, r)$ and a permutation
  $\sigma\in R_k$, $\alpha (q\sigma) = \alpha(q)\sigma$,  where
  $\sigma $ acts on the function $\alpha(q)$ on the right by
  permuting its arguments.
\end{enumerate}
\item we introduce the two functions, $\partial_0 ~ and~ \partial_1$,
such that for any  $\alpha(q)$,  $~q\in O(r_1,..., r_k, r)$, we have
that
   $\partial_0(q) = \{r_1,..., r_k\}$,
   $~\partial_0(\alpha(q)) = \{\alpha(r_1),..., \alpha(r_k)\}$,
   $~\partial_1(q)= \{r\}$, and $\partial_1(\alpha(q))= \{\alpha(r)\}$.
\end{enumerate}
\end{definition}
Consequently, we can think of an operad as a simple sort of theory,
used to define a schema mappings between databases, and its algebras
as models of this theory used to define the mappings between
instance-databases, where a mapping $\alpha$ is considered as an
interpretation of relation symbols of  a given database schema.
%
\section{Data Object Type for query-answering database systems}
We consider the views as a universal property for databases:
  they are the possible observations of the information contained in
  an  instance-database, and we can use them in order to establish an equivalence relation
  between databases.\\
In a theory of \emph{algebraic specifications} an Abstract Data Type
(ADT) is specified by a set of operations (constructors) that
determine how the values of the carrier set are built up, and by a
set of formulae (in the simplest case, equations) stating which
values should be identified. In the standard initial semantics, the
defining equations impose a congruence on the initial algebra.
Dually, a \emph{coagebraic specification} of a class of systems,
i.e., Abstract Object Types (AOT), is characterized by a set of
operations (destructors) that specify what can be \emph{observed}
out of a system-\emph{state} (i.e., an element of the carrier), and
how  a state can be transformed to a
successor-state.\\
We start by introducing the class of coalgebras for database
query-answering systems for a given instance-database (a set of
relations) $A$. They are presented in an algebraic style, by
providing a co-signature. In particular, the sorts include one
single "hidden sort" corresponding to the carrier of the coalgebra,
and other "visible" sorts, for inputs and outputs, that have  a
given  fixed interpretation. Visible sorts will be interpreted as
sets without any algebraic structure defined on them. For us,
coalgebraic terms, built only over destructors, are precisely
interpreted  as the basic \emph{observations} that one can make on
the states of a coalgebra. \\
Input sorts are considered as a set $\L_A$ of union of conjunctive
queries $q(\textbf{x})$ for a given database $A$, where $\textbf{x}$
is a tuple of variables (attributes) of this query. Each query has
an algebraic term of the "select-project-join + union" algebraic
query language (SPJRU, or equivalent to it, SPCU algebra, Chapter
4.5, 5.4 in ~\cite{AbHV95}) with a carrier equal to the set of
relations in $A$.
 We define  the power
  view-operator $T$, with domain and codomain equal to the set of all
  instance-databases, such that for any object (database) $A$, the object
  $TA$  denotes a database composed by the set of \emph{all views} of $A$.
   The object $TA$, for a given instance-database $A$, corresponds to
   the  quotient-term  algebra $\L_A/_\approx$,
  where the carrier is a set of equivalence classes of closed terms of a
well-defined formulae of a relational algebra. Such formulae are
"constructed" by $\Sigma_R$-constructors (relational operators in
SPJRU algebra: select, project, join and union), by symbols
(attributes of relations) of a database instance $A$, and by
constants of attribute-domains.
  More precisely, $TA$ is "generated" by this quotient-term algebra $\L_A/_\approx$.
   For every object $A$ holds that $A \subseteq TA$, and $TA = TTA$, i.e., each (element) view
   of database instance $TA$ is also an element (view) of a database instance
   $A$.
   Notice that  when $A$ is also  finitary
   (has a finite number of relations) but
   with at least one relation with infinite number of tuples, then $TA$ has an infinite number of relations (views of $A$),
    thus can be an infinitary object.
   It is obvious that when a domain of constants of a database is finite then both $A$ and $TA$ are finitary objects. As
   default we assume that a domain of every database is an arbitrary  large finite set. This is a reasonable assumption for  real applications.\\
Consequently, the output sort of this database AOT is a set  $TA$ of
all resulting views (resulting n-ary relation) obtained by
computation of queries $q(\textbf{x})$ over a database $A$. It is
considered as the carrier of a coalgebra as well.
\begin{definition} A co-signature for a Database query-answering system, for a given instance-database A, is
a triple $\D_\Sigma = (S, OP, [\_~])$, where S are the sorts,  OP
are the operators, and [\_~] is an interpretation of visible sorts,  such that: \\
1. $S = (X_A, \L_A, \Upsilon)$, where $X_A$ is a hidden sort (a set
of states of a database A), $\L_A$ is an input sort (set of union of
conjunctive queries), and $\Upsilon$ is an output sort (the set of
all views of of all instance-databases).\\
2. OP is a set of operations: a method $~~Next:X_A\times \L_A
\rightarrow X_A $, that corresponds to an execution of a next query
$q(\textbf{x}) \in \L_A$ in a current state of a database A, such
that a database A passes to the next state; and $~~Out:X_A\times
\L_A \rightarrow TA$ is an attribute that returns with the obtained
view of a database for a given query $~q(\textbf{x})\in \L_A$.\\
3. [\_~] is a function, mapping each visible sort to a non-empty
set.\\
The Data Object Type for a query-answering system is given by a
coalgebra:\\ $~~< \lambda Next, \lambda Out >:X_A \rightarrow
X_A^{\L_A} \times  TA^{\L_A}~$, of the polynomial endofunctor
$~~(\_~)^{\L_A} \times  TA^{\L_A}:Set\rightarrow Set$, where
$\lambda$ denotes the lambda abstraction for functions of two
variables into functions of one variable (here $Z^Y$ denotes the set
of all functions from Y to Z).
\end{definition}
This separation between the sorts and their interpretations is given
in order to obtain a conceptual clarity: we will simply ignore it in
the following by denoting both, a sort and the corresponding set, by
the same symbol. In an object-oriented terminology, the coalgebras
are expressive enough in order to specify the parametric methods and
the attributes for a database (conjunctive) query answering systems.
In a transition system terminology, such coalgebras can model a
deterministic, non-terminating, transition system with inputs and
outputs. In ~\cite{Corr97}  a complete equational calculus for such
coalgebras of restricted class of polynomial functors has been defined. \\
 In the
rest of this paper we will consider only the database
query-answering systems without side effects: that is, the obtained
results (views) \emph{will not be materialized} as a new relation of
this database $A$. Thus, when a database answers  a query, it
remains in the same initial state. Thus, the set $X_A$ is a
singleton $\{ A \}$ for a given database $A$, and consequently it is
isomorphic to the terminal object $1$ in the $Set$ category. As a
consequence, from $1^{\L_A} \simeq 1$, we obtain that a method
$Next$ is just an identity function $id:1 \rightarrow 1$.
Consequently, the only interesting part of this AOT, is the
attribute part $~~Out:X_A \times \L_A \rightarrow TA$, with the fact
that $X_A \times \L_A = \{A\} \times \L_A \simeq \L_A$.
\\Consequently, we obtain an attribute mapping $~~Out:\L_A \rightarrow TA$,
which will be used as a semantic foundation for a definition of
database mappings:  for any query $q_i(\textbf{x})\in \L_A$, the
corespondent algebraic term $~\widehat{q_i}$ is a function (it is
not a T-coalgebra) $\widehat{q_i}:A^k \rightarrow TA$, where $A^k$
is k-th cartesian product of $A$ and $r_{i1},...,r_{ik} \in A$ are
the relations used for computation of this query. A view-mapping can
 be defined now as a \emph{T-coalgebra} $q_{A_i}:A \rightarrow TA$,
that, obviously, \emph{is not} a function. We introduce also the two
functions $\partial_0, \partial_1$ such that $\partial_0(q_{A_i}) =
\{r_{i1},...,r_{ik}\}$ and $\partial_1(q_{A_i}) = \{r_i\}$, with
obtained  view $ r_i = \|q_i(\textbf{x})\| = \widehat{q_i}
(r_{i1},...,r_{ik})$. Thus, we can formally introduce a theory for
operads:
\begin{definition}\label{def:view-map} \textsc{View-mapping:}
  For any \textsl{query} over a  schema $\A$  we can define a
  schema map $q_i:\A\longrightarrow T\A $,  where $q_i\in O(r_{i1},..., r_{ik}, r_i )$,
  $~Q = (r_{i1},..., r_{ik})\subseteq~\A$,  and $r_i\in T\A$.\\  A correspondent view-map
  at instance level is $q_{A_i}= \{\alpha(q_i), q_\perp\}:A \longrightarrow TA$, with $A = \alpha^*(\A)$,
  $TA = \alpha^*(T\A)$, $~\partial_0(q_\perp)= \partial_1(q_\perp)=
  \{\perp\}$. For simplicity, in the rest of this paper we will drop the
  component $q_\perp~$ of a view-map, and assume implicitly  such a
  component; thus,
    $~\partial_0(q_{A_i}) = \alpha^*(Q)\subseteq A ~$ and $~\partial_1(q_{A_i})= \{\alpha(r)\} \subseteq TA
  ~$ is a singleton with the unique element equal to view obtained by a "select-project-join+union"
  term $~\widehat{q_i}$.
\end{definition}

\section{Database category DB}
\begin{figure}
 $\vspace*{-12mm}$
\begin{center}
 \includegraphics{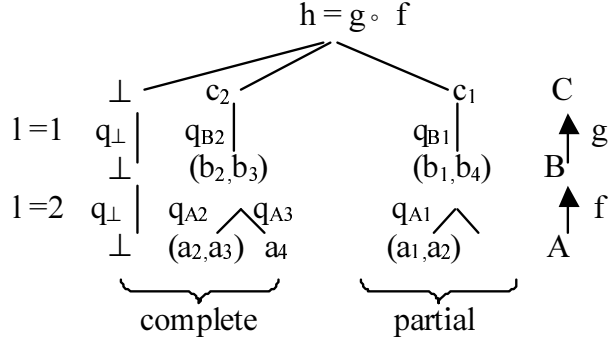}
 \caption{obtained partial morphism}
 \label{partial-morphism}
 \end{center}
 $\vspace*{-9mm}$
 \end{figure}

\begin{figure}
  $\vspace*{-9mm}$
\begin{center}
 \includegraphics{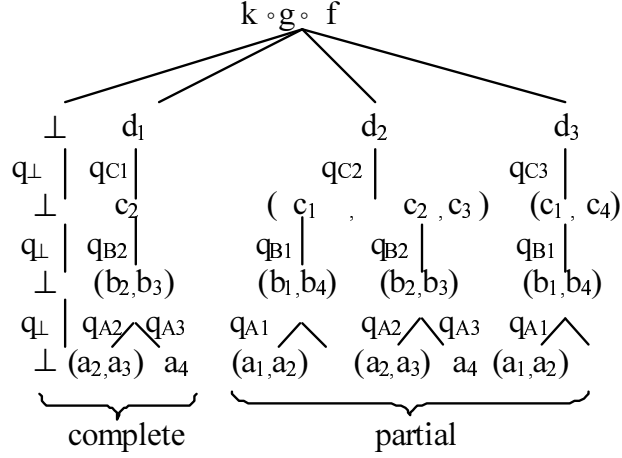}
 \caption{composed morphism}
 \label{composed-morphism}
 \end{center}
 $\vspace*{-10mm}$
\end{figure}

Based on an observational point of view for relational databases, we
may introduce a category $DB$ \cite{Majk04AOT} for
instance-databases and view-based mappings between them, with the
set of its objects $Ob_{DB}$, and the set of its morphisms
$Mor_{DB}$, such that:
\begin{enumerate}
  \item Every object  (denoted by $A,B,C$,..) of this category is a
  instance-database, composed by a set of n-ary relations  $a_i\in A$, $i= 1,2,...$  called also "elements of
  $A$".
 We define a universal database instance  $\Upsilon$
  as the  union of all database instances, i.e., $\Upsilon = \{ a_i | a_i\in A, A\in Ob_{DB}\}$.
  It  is the top object of this category.\\
 A \emph{closed object} in $DB$ is a instance-database $A$ such that $A = TA$.
     We have that $\Upsilon = T\Upsilon$, because every view $v\in T\Upsilon$
  is  an instance-database as well, thus $v\in \Upsilon$. Vice versa, every element
   $r\in \Upsilon$ is  a view of $\Upsilon$ as well, thus $r\in T\Upsilon$.\\
      Every object (instance-database) $A$ has also an empty relation $\bot$. The object  composed by only this
  empty relation is denoted by $\bot^0$ and we have that $T\bot^0
  =\bot^0= \{\bot\}$. Any empty database (a database with only empty relations) is isomorphic to this bottom object $\bot^0$.
  \item Morphisms of this category are all possible mappings
  between instance-databases \emph{based on views}, as they will be defined
  by formalism of operads in what follows.
\end{enumerate}
In what follows, the objects in $DB$ (i.e., instance-databases) will
be called simply databases as well, when it is clear from the
context.
Each atomic mapping (morphism) in $DB$ between two databases is
generally composed of three components: the first correspond to
conjunctive query $q_i$ over a source database that defines this
view-based mapping, the second (optional) $w_i$ "translate" the
obtained tuples from domain of the source database (for example in
Italian) into terms of domain of the target database (for example in
English), and the last component $v_i$ defines which contribution of
this mappings is given to the target relation, i.e., a kind of
Global-or-Local-As-View (GLAV) mapping
(sound, complete or exact).\\
Instead of lists $(g_1,...,g_k)$ used
  for mappings in Definitions \ref{def:operads}, \ref{def:operads2}, we will use the sets
  $\{g_1,...,g_k\}$ because a mapping between two databases does not
  depend on a particular permutation of its components.
Thus, we introduce an \emph{atomic morphism} (mapping) between two
databases as a set of simple view-mappings:
\begin{definition}\label{def:atomicmorphisms} \textsc{Atomic morphism:}
Every \textsl{schema mapping} $~f_{Sch}:\A\longrightarrow \B$, based
on a set of query-mappings $q_i$, is defined
  for finite natural number N by\\
  $ f_{Sch} \triangleq \{~v_i\cdot w_i\cdot q_i~ |~ q_i\in O(r_{i1},..., r_{ik},~
  r_i''), ~w_i\in O(r_i'',r_i'), ~v_i\in O(r_i',r_i),\\~ \{r_{i1},..., r_{ik}\}\subseteq \A ,~ r_i\in
  \B, 1\leq i\leq N \}$.\\
Its correspondent \textsl{complete morphism} at instance database level is\\
  $f = \alpha^*(f_{Sch}) \triangleq \{~q_{A_i} = \alpha(v_i)\cdot \alpha(w_i)\cdot \alpha(q_i)~|~
  ~ v_i\cdot w_i\cdot q_i \in f_{Sch}\}:A \rightarrow B$,
where:\\
  Each $\alpha (q_i)$ is a query computation, with obtained view $\alpha(r_i'') \in TA$
  for an instance-database  $A = \alpha^*(\A) = \{ \alpha(r_k)~|~r_k \in \A \}$, and $B = \alpha^*(\B)$.\\
  Each $\alpha(w_i):\alpha(r_i'')\longrightarrow \alpha(r_i')$, where $\alpha(r_i') \in TB$, is
 equal to the function determined by the symmetric \textsl{domain
 relation} $~r_{AB}\subseteq \textbf{dom}_A \times \textbf{dom}_B~$
 for the equivalent constants in $\alpha^*(\A)~and~\alpha^*(\B)~$ ($(a,b)\in r_{AB}~$ means that, $a \in \textbf{dom}_A~and~b \in
 \textbf{dom}_B~$ represent the same entity of the real word (requested for a federated database environment) as:
 for any $(a_1,...,a_n) \in \alpha(r'')~holds~
 \alpha(w_i)(a_1,...,a_n)=(b_1,...,b_n)$, and for all $1\leq
 k\leq n ~~(a_k,b_k) \in r_{AB}$.
 If $r_{AB}~$ is not defined, it is assumed, by default,  that
 $\alpha(w_i)~$ is an identity function.\\
 Let $P_{q_i}$ be a projection function on relations, for all attributes in
 $\partial_1(\alpha(q_i)) = \{\alpha(r_i'')\}$. Then,
  each $\alpha (v_i):\alpha(r_i')\longrightarrow \alpha (r_i)$ is one  tuple-mapping function,
   used to distinguish sound,
   complete and exact assumptions on the views, as follows:
\begin{enumerate}
  \item \textsl{inclusion} case, when $~ \alpha(r_i')\subseteq P_{q_i}(\alpha
  (r_i))$. Then
    for any tuple $t\in \alpha(r_i')$, $~~\alpha (v_i)(t)
  = t_1$, for some $t_1\in\alpha(r_i)~$ such that $P_{q_i}(\{t_1\})=
  t$.\\ We define  $\|q_{A_i} \| \triangleq  \alpha(r_i')$ the extension of data
  transmitted from an instance-database $A$ into $B$ by a component $q_{A_i}$.
  \item \textsl{inverse-inclusion} case, when $ \alpha(r_i')\supseteq P_{q_i}(\alpha
  (r_i))$.\\
  Then, for any tuple $t\in \alpha(r_i')$,
  \begin{displaymath}
   \alpha (v_i)(t) = \left\{ \begin{array}{ll}
   t_1 & \textrm{ ~, ~if ~$ \exists{t_1}\in \alpha (r_i)~,~P_{q_i}(\{t_1\})
  = t$}\\
& \textrm{~empty ~tuple, ~otherwise}
  \end{array} \right.
  \end{displaymath}
  We define  $\|q_{A_i} \| \triangleq  P_{q_i}(\alpha
  (r_i))$ the extension of data
  transmitted from an instance-database  $A$ into $B$ by a component $q_{A_i}$.
  \item \textsl{equal} case, when both (a) and (b) are valid.
  \end{enumerate}
     \end{definition}
  Notice that the components $\alpha(v_i), \alpha(w_i), \alpha(q_i)$ are
  not the morphisms in $DB$ category:  only their functional
  composition is an atomic morphism.
  Each atomic morphism is a complete morphism, that is, a set of view-mappings. Thus,
  each view-map $~q_{A_i}:A\longrightarrow TA$, which is an atomic morphism,
   is a complete morphism (the case when $B = TA$, $r_{AB}~$ is not defined, and
   $\alpha (v_i)$ belongs to the "equal case"),
  and  by \emph{c-arrow} we denote the set of all complete  morphisms.\\
 \textbf{Example 1}: In the Local-as-View (LAV) mappings ~\cite{Lenz02}, the inverse
inclusion, inclusion and equal case correspond to the sound ,
complete and exact view respectively. In the Global-as-View (GAV)
mappings, the inverse inclusion, inclusion and equal case correspond
to the complete,
sound and exact view respectively.\\$\square$\\
 \textbf{Remark:} In
the rest of this paper we will consider only empty domain relations
(i.e., when $\alpha(w_i)$ are the identity functions) and we will
write $r\in A$ also for $\alpha (r)\in \alpha^*(\A)$, i.e., the name
(type) of a relation $r$ in $\A$ is  used also for its extension
(set of tuples of that relation), and $A$  for $\alpha^*(\A)$ as
well. Notice that the functions $\partial_0 ~ and~
\partial_1$ are different from $dom$ and $cod$ functions used for
the category arrows. Here $\partial_0$ specifies exactly the subset
of relations in a database $A$ used for view-based mapping, while
$\partial_1$ defines the target relation in a database $B$ for this
mapping. Thus: $\partial_0(f) \subseteq~ dom(f) = A$, $~~
\partial_1 (f) \subseteq ~cod(f) = B$ (in the case when $f$
 is a simple view-mapping then $\partial_1 (f)$ is a singleton).
In fact, we have that they are functions $~\partial_0,
\partial_1:Mor_{DB} \rightarrow \P(\Upsilon)$ (where $\P$ is the powerset
operation), such that for any morphism $f:A \rightarrow B$ between
databases $A$ and $B$, we
have that $\partial_0(f) \subseteq A$ and $\partial_1(f) \subseteq  B$.\\
The \emph{Yes/No query} $q_i$ over a database $A$, obviously do not
transfer any information to target object $TA$. Thus, if the answer
to such a query is $Yes$, then this query is represented in $DB$
category as a mapping $q_i:A \rightarrow TA$, such that the source
relations in $\partial_0(q_i)$ are non-empty and $\partial_1(q_i) =
\{\bot\}$. The answer to such a query $q_i$ is $No$ iff (if and only
if)  such a mapping does not exist
in this $DB$ category.\\$\square$\\
 We are ready now to give a
formal definition for \emph{all}  morphisms in the category $DB$.
Generally, a composed  morphism $h:A \rightarrow C$
 is a general tree such that all its leaves are
 not in $A$: such a morphism is denominated as an \emph{incomplete} (or partial) p-arrow.
\begin{definition}\label{def:morphisms} \textsc{Sintax}: The following BNF defines the set  of all morphisms in
DB:\\
  $~~~p-arrow \textbf~{:=} ~c-arrow ~|~c-arrow \circ c-arrow~$ (for any two c-arrows  $f:A\longrightarrow B$ and $g:B\longrightarrow C~$)\\
 $~~~morphism \textbf~{:=} ~p-arrow ~|~c-arrow \circ p-arrow~$ (for any p-arrow $f:A\longrightarrow B$ and c-arrow $g:B\longrightarrow C$)\\
\\whereby the composition of two arrows, f (incomplete) and g (complete),
we obtain the following p-arrow  $h = g \circ f:A\longrightarrow C$
\[
\ h = g\circ f = \bigcup_{ q_{B_j}\in ~\alpha^*(g_{Sch}) ~\&
~\partial_0 (q_{B_j}) \bigcap
\partial_1 (f) \neq \emptyset } \{q_{B_j}\}~~~~\circ
\]
\[
\circ ~~~~ \bigcup_{q_{A_i}\in ~\alpha^*(f_{Sch}) ~\&~\partial_1
(q_{A_i})= \{v\}~\&~ v \in ~\partial_0 (q_{B_j}) }
\{q_{A_i}(tree)\}~~
\]
$= \{q_{B_j} \circ \{q_{A_i}(tree)~|~\partial_1(q_{A_i}) \subseteq
\partial_0(q_{B_j})\}~|~q_{B_j}\in ~\alpha^*(g_{Sch}) ~\&
~\partial_0 (q_{B_j}) \bigcap
\partial_1 (f) \neq \emptyset \}$\\
$= \{q_{B_j}(tree)~|~q_{B_j}\in ~\alpha^*(g_{Sch}) ~\& ~\partial_0
(q_{B_j}) \bigcap
\partial_1 (f) \neq \emptyset\}$
\\\\where   $q_{A_i}(tree)$ is the  tree of the morphisms f below
$q_{A_i}$.
\end{definition}
 We have the equal analog diagrams of \emph{schema}
mappings as well:
\begin{itemize}
  \item For a morphism $f:A\longrightarrow B~$ in DB we have
  syntactically identical schema mapping arrow $f_{Sch}:\A\longrightarrow
  \B~$ without the interpretation of its symbols (the composition of
  functions $" \circ "$ is replaced by the associative composition
  of operads $" \cdot "$)
  \item A \emph{schema mapping graph} G is any subset of schema arrows.
\end{itemize}
\begin{figure}
 $\vspace*{-12mm}$
\begin{center}
 \includegraphics{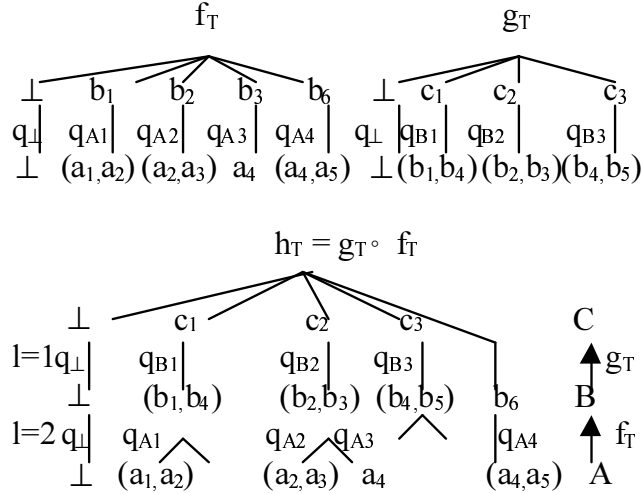}
 \caption{composed tree}
 \label{composed-tree}
 \end{center}
 $\vspace*{-9mm}$
 \end{figure}
Notice that the arrows (morphisms) in $DB$ are not functions. Thus,
$DB$ is different from  $Set$ category.
 In order to explain the composition of  morphisms let us consider the following example: \\
 \textbf{Example 2}: Let us consider the morphisms $f:A\longrightarrow
 B$, $g:B\longrightarrow C$, such that\\
 $ A = \{a_1,..,a_6\},~B =\{b_1,..,b_7\},~C
 =\{c_1,..,c_4\},~where~ f =\{q_{A_1},..,q_{A_4}\},
 with~\\ \partial_0(q_{A_1})=\{a_1,a_2\},~\partial_0(q_{A_2})=\{a_2,a_3\},
 ~\partial_0(q_{A_3})=\{a_4\},~\partial_0(q_{A_4})=\{a_4,a_5\},\\~
 \partial_1(q_{A_1})= \{b_1\},~\partial_1(q_{A_2})= \{b_2\},
 ~\partial_1(q_{A_3})= \{b_3\},~\partial_1(q_{A_4})= \{b_6\},~
~and~ g =\{q_{B_1},..,q_{B_3}\},\\
 with~ \partial_0(q_{B_1})=\{b_1,b_4\},~\partial_0(q_{B_2})=\{b_2,b_3\},
 ~\partial_0(q_{B_3})=\{b_4,b_5\},~\partial_1(q_{B_1})= \{c_1\},~\partial_1(q_{B_2})= \{c_2\},
 ~\partial_1(q_{B_3})= \{c_3\}$, that can be represented by
 trees $f_T = f$ and $g_T = g$ and their sequential composition $h_T$ (Fig.
 \ref{composed-tree}).\\
The composition of morphisms (Fig. \ref{partial-morphism}) $h =
g\circ f:A\longrightarrow C$ may be represented as \emph{a part of
the tree} $h_T$  that gives information contribution from the object
$A$ (source) into the object $C$ (target of this composed morphism).
We have that $\partial_0(f) = \{a_1,a_2,a_3,a_4,a_5\}$,
$\partial_1(f) = \{b_1,b_2,b_3,b_6\}$,
$\partial_0(g) = \{b_1,b_2,b_3,b_4,b_5\}$,
$\partial_1(g) = \{c_1,c_2,c_3\}$, while $\partial_0(h)
=\partial_0(g\circ f) = \{a_1,a_2,a_3,a_4\}\neq
\partial_0(f),~~\partial_1(h)
=\partial_1(g\circ f) = \{c_1,c_2\}\neq \partial_1(g)$\\
  Let us see, for example, the composition of the c-arrow
$h:C\longrightarrow D$ with the composed arrow $g\circ f$ in the
previous example, where $D = \{d_1,..,d_4)\}$, $h =
\{q_{C_1},q_{C_2},q_{C_3}\}$,
$~\partial_0(q_{C_1})=\{c_2\}$,$~\partial_1(q_{C_1})=\{d_1\}$,
$~\partial_0(q_{C_2})=\{c_1,c_2,c_3\}$,$~\partial_1(q_{C_2})=\{d_2\}$,
$~\partial_0(q_{C_3})=\{c_1,c_4\}$,$~\partial_1(q_{C_3})=\{d_3\}$,
with
$q_{B_2}(tree)= q_{B_2}\circ \{q_{A_2},~q_{A_3}\}$ a complete, and
$q_{B_1}(tree)= q_{B_1}\circ \{q_{A_1},~-~\}$ a partial (incomplete)
component of this tree, as
represented in the Fig. \ref{composed-morphism}.\\
 As we see, a composition of (complete) morphisms
generally produces a partial (incomplete) morphism (only a part of
the tree $h_T$ represents a real contribution from $A$ into $C$)
with \emph{hidden elements} (in the diagram of the composed morphism
$h$, the element $b_4$ is a hidden element). In such a
representation we "forgot" parts of the tree $g_T\circ f_T$ that are
not involved in real information contribution of composed mappings
from the source into the target object. So, we define the semantics
of any morphism $h:A\longrightarrow C$  as an "information
 transmitted flux"   from the source into the target
object. An "information flux" (denoted by $~\widetilde{h}$) is a set
 set of views (so, it is an object in $DB$ category as well) which is
"transmitted" by a mapping.
\\$\square$

In order to explain this concept of "information flux" let us
consider a simple morphism $f:A\longrightarrow B~$ from a database
$A$ into a database $B$, composed by only one view map based on a
single query $q(\textbf{x})\longleftarrow R_1(u_1), ..., R_n(u_n)$,
where $n\geq 0$, $R_i$ are relation names (at least one) in $\A$ or
built-in predicates (ex. $\leq, =,$ etc..), and $q$ is a relation
name not in $\A$. Then, for any tuple $\textbf{c}$ for which the
body of this query is true, also $q(\textbf{c})$ must be true, that
is, this tuple from a database $A$ "is transmitted" by this
view-mapping into one relation of database $B$. The set (n-ary
relation) $Q$ of all tuples that satisfy the body of this query will
constitute the whole information "transmitted" by this mapping. The
"information flux" $ \widetilde{f}$ of this mapping is  the set
$TQ$, that is, the set of all views (possible observations) that can
be obtained from the transmitted information of this mapping.
\begin{definition} \label{semantix-morphism} We define the semantics of mappings by function
 $B_T:Mor_{DB}\longrightarrow Ob_{DB}$, which, given any mapping
 morphism
 $f:A\longrightarrow B~$, returns with the set of views ("information flux") that are
 really "transmitted" from the source to the target object.\\
 1. For an atomic morphism, $ \widetilde{f} = B_T(f)~\triangleq
 T\{\|q_{A_i}\|~|~q_{A_i} \in f \}$.\\
 2. Let $g:A \rightarrow B$ be a morphism with a flux
 $\widetilde{g}$, and $f:B \rightarrow C$ an atomic morphism with
 flux $\widetilde{f}$ defined in point 1, then  $~~\widetilde{f \circ g}
 = B_T(f \circ g) ~\triangleq \widetilde{f}\bigcap
 \widetilde{g}$.
 \end{definition}
 Thus we have the following fundamental property:
\begin{propo}\label{prop-morphism} Any mapping  morphism
 $f:A\longrightarrow B~$  is a closed object in DB, i.e., $~~\widetilde{f} =
T\widetilde{f}$.
\end{propo}
\textbf{Proof: }
This proposition may be proved by  structural induction; each atomic
arrow is a closed object ($T\widetilde{f} =
T(T\{\|q_{A_i}\|~|~q_{A_i} \in f \}) = T\{\|q_{A_i}\|~|~q_{A_i} \in
f \} = \widetilde{f}$, each arrow is a composition of a number of
complete arrows, and intersection of closed objects is always a
closed object.
\\$\square$\\
\textbf{Remark: }The "information flux" $\widetilde{f}$ of a given
morphism (mapping) $f:A\longrightarrow B$ is an instance-database
 as well (its elements are the views defined by the formulae
above), thus, an object in $DB$: the minimal "information flux" is
equal to the bottom object $\bot^0$ so that, given any two database
instances $A,B$ in $DB$, there exists at least an arrow (morphism)
between them $f:A\longrightarrow B$ such that $\widetilde{f} =
\bot^0$.
\begin{propo} \label{prop:morphisms} The following properties for morphisms
are valid:
\begin{enumerate}
  \item each arrow $f:A\twoheadrightarrow B$, such that $~\widetilde{f} =
  TB$ is an epimorphism
  \item each arrow $f:A\hookrightarrow B$, such that $~\widetilde{f}=
  TA$ is a monomorphism
  \item each monic and epic arrow is an isomorphism, thus
  two objects A and B are isomorphic iff $~~TA = TB$, i.e., \[
  A \backsimeq B ~~~~iff~~~~ TA = TB \]
\end{enumerate}
\end{propo}
\textbf{Proof:} 1. An arrow $f:A\twoheadrightarrow B$ is epic iff
for any $h,g:B\longrightarrow C$ holds $(h\circ f = g\circ f)
\Rightarrow (h = g)$, thus $(\widetilde{h \circ f} = \widetilde{g
\circ f}) \Rightarrow(\widetilde{h} = \widetilde{g})$
which is satisfied by $\widetilde{f} = TB$ (because $\widetilde{h}\subseteq TB ~and~ \widetilde{g}\subseteq TB$)\\
2. An arrow $f:A\longrightarrow B$ is monic iff  for any
$h,g:C\longrightarrow A$ holds $(f\circ h = f\circ g) \Rightarrow(h
= g)$, thus $(\widetilde{f \circ h} = \widetilde{f \circ g})
\Rightarrow (\widetilde{h} = \widetilde{g})$
which is satisfied by $\widetilde{f} = TA$ (because $\widetilde{h}\subseteq TA ~and~ \widetilde{g}\subseteq TA$)\\
3. By 1 and 2, because an isomorphism is epic and monic, and
viceversa if $f$ is monic and epic then $\widetilde{f} = TA$ (2) and
$\widetilde{f} = TB$ (1), thus $TA = TB$. It is enough to show the
isomorphism $A \backsimeq TA$ : let us define the isomorphisms
$is_A:A\longrightarrow TA$, and its inverse
$is^{-1}_A:TA\longrightarrow A$,
\[
\ is_A=\bigcup_{ \partial_1(q_{A_i})=\{v\} ~\& ~v \in  TA }~
\{q_{A_i}\},~~~~ is^{-1}_A=\bigcup_{ \partial_0(q_{TA_i}) =
\partial_1(q_{TA_i}) ~\& ~\partial_1(q_{TA_i}) =\{v\} ~\& ~v \in A }~
\{q_{TA_i}\}
\]
 Thus, $\widetilde{is_A} =
\widetilde{is^{-1}_A} = TA$, so it holds that
$\widetilde{is^{-1}_A\circ is_A} = TA = \widetilde{id_A} =
\widetilde{is_A\circ is^{-1}_A}$, i.e.,  $is^{-1}_A\circ is_A =
id_A$ and $is_A\circ is^{-1}_A = id_{TA}$, thus $A \backsimeq TA$.
Finally, $~~A \backsimeq TA = TB \backsimeq B$, i.e.,$~~A \backsimeq
B$.
\\$\square$\\
\textbf{Remark:}
 Thus, we consider, for example, the real object (empty database instance) $\bot^0$ as \emph{zero object} (both terminal and initial) in $DB$,
(from any real object $A$ in $DB$ there is a unique arrow from it
into $ \bot^0$ and its reversed arrow). Each arrow $f$ with
$\partial_0(f) = \{\bot\}$ or $\partial_1(f) = \bot$  has an empty
flux, thus does not give any information contribution to the target
database: as for
example \emph{Yes arrows} in $DB$ for Yes/No queries.\\
It is easy to verify that each \emph{empty database}  (with all
empty relations) is isomorphic to the zero object $\bot^0$.\\
In what follows we will show that any two isomorphic objects
(databases) in $DB$ are observationally equivalent.
\subsection{Interpretations of schema mappings}
  The semantics of  mapping between two relational database schemas, $f:\A\longrightarrow \B$,  is a
  constraint on the pairs of interpretations, of $\A$ and $\B$,
  and therefore specifies which pairs of interpretations can
  co-exist, given the mapping (see also ~\cite{MBDH02}).
  We consider only view-based mappings between schemas defined in the SQL language of $SPJRU$
  algebra, i.e.,   when \\
   $(1)~~~f = \{q_{Ai}(x) \Rightarrow b_j(x)\}$, where
  $q_{Ai}(x)$ is a union of conjunctive queries  over $\A$ and $b_j$ is a relation symbol of a database
  schema $\B$, or,\\
 $(2)~~~f = \{q_{Ai}(x) \Rightarrow q_{Bj}(x)\}$, where
  $q_{Bj}(x)$ is a union of conjunctive queries over $\B$.  In this case the mapping $f$
  also involves a helper database schema $\C$ with a relation $c_i(x)$ for each $q_{Ai}(x) \in f$ with  two new
  database mappings, $f_{AC}:\A\rightarrow \C$ and $f_{BC}:\B\rightarrow \C$, with
   $f_{AC} = \{q_{Ai}(x) \Rightarrow c_i(x)\}$ and $f_{BC} = \{q_{Bj}(x) \Rightarrow c_i(x)\}$.\\
  The  formula $e = ~q_{Ai}(x) \Rightarrow
  q_{Bj}(x)$ (logical implication between queries), means that each tuple of the view obtained by the
  query $q_{Ai}(x)$ is also a tuple of the view obtained by the
  query $q_{Bj}(x)$.\\
 There is a fundamental functorial \emph{interpretation} connection from schema
 mappings and their models in the instance level category $DB$:
 based on the Lawvere categorial theories \cite{Lawve63,BaWe85}, where he
 introduced a way of describing algebraic structures using
 categories for theories, functors (into base category $Set$, which
 we will substitute by more adequate category $DB$), and natural
 transformations for morphisms between models. For example, Lawvere's seminal
 observation that the theory of groups is a category
 with group object, that group in $Set$ is a product preserving
 functor, and that a morphism of groups is a natural transformation
 of functors, is an original new idea that was successively
 extended in order to define the categorial semantics for different
 algebraic and logic theories. This work is based on the theory of
 \emph{sketches}, which are fundamentally graphs enriched by other
 concepts such as (co)cones mapped by functors in (co)limits of the base
 category $Set$. It was demonstrated that, for every sentence in
 basic logic, there is a sketch with the same category of models, and
 vice versa \cite{Mapa89}. Accordingly, sketches are called
 graph-based logic and provide very clear and intuitive
 specification of computational data and activities. For any small sketch
 $E$ the category of models $Mod(E)$ is an accessible category by Lair's theorem and
 reflexive subcategory of $Set^E$ by Ehresmann-Kennison theorem. In
 what follows we will substitute the base category $Set$ by this new
 database category $DB$.
 \begin{propo} \label{prop:schema}
 Let $Sch(G)~$ be a schema category generated from a schema
 mapping graph (sketch) $G$ . Every interpretation R-algebra $\alpha~$ has
 as its categorial correspondent the functor (categorial model) $~\alpha^*:Sch(G)\longrightarrow
 DB~$, defined as follows:
\begin{enumerate}
  \item for any database schema $\A = \{a_1,...,a_n\}$, (object in $Sch(G)$),
  where $ a_i\in R,~ i=1,..,n $, holds $A \triangleq \alpha^*(\A) =
  \{\alpha(a_1),...,\alpha(a_n)\}$, i.e., $A$ is an interpretation (logical model) of a database schema $\A$.
  \item for any schema mapping arrow $f:\A\longrightarrow \B$, let $~~f_T~~$ be the tree structure of operads,
   $ f_T = \{f_1 \cdot g_1,...,f_k \cdot g_k)\}$, where each $f_i $
   is a linear composition of operads, then $\alpha^* (f) =
   \{\alpha(f_1)\circ \alpha^*( g_1),...,\alpha(f_k)\circ\alpha^*(g_k)\}$, otherwise $\alpha^*
   (f) = \alpha (f_T)$.\\
   Formally, the satisfaction of mapping $f$ is defined as follows: for each logical formula $~~e \in f$, $~~~\{ \alpha^*(\A),
   \alpha^*(\B)\}\vDash ~e$, that is $e$ is \verb"satisfied" by a model
   $\alpha^* \in Mod(Sch(G)) \subseteq DB^{Sch(G)}$.
    \end{enumerate}
 \end{propo}
 \textbf{Proof:} This is easy to verify, based on general theory for sketches \cite{BaWe85}: each arrow in a sketch  (enriched schema mapping graph) $G$ may be
 converted into a tree syntax structure of some morphism in $DB$ (labeled tree without
 any interpretation), thus, a sketch $G$ can be extended into a
 category $Sch(G)$.
 (The composition of schema mappings in the category $Sch(G)$,
 where each mapping is a set of first-order logical formulas,
 can be defined as a disjoint union). The functor is
 only the simple extension of the interpretation R-algebra function
 $\alpha~$ for a lists of symbols, as in Definition \ref{def:atomicmorphisms}.
\\$\square$
%
\subsection{Power-view endofunctor T}
 Let us extend the notion of the type operator $T$ into a
notion of the endofunctor in $DB$ category:
\begin{theo} \label{th:endofunctor}  There exists an endofunctor $T =
(T^0,T^1):DB \longrightarrow DB$, such that
\begin{enumerate}
  \item for any object A, the object component $T^0$ is
  equal to the type operator T, i.e.,  $~~~~T^0(A)\triangleq TA$
  \item for any morphism $~f:A\longrightarrow B$, the arrow
  component $T^1$ is  defined by
  \[
  \ T(f) \triangleq T^1(f) =  \bigcup_{ \partial_0(q_{TA_i})
  =\partial_1(q_{TA_i}) =\{v\} ~\& ~v \in ~ \widetilde{f}} \{q_{TA_i}:TA
  \rightarrow TB\}
  \]
  \item Endofunctor T preserves the properties of arrows, i.e., if a
  morphism $f$ has a property P (monic, epic, isomorphic), then
  also $T(f)$ has the same property: let $P_{mono}, P_{epi}
  ~and~\\P_{iso}$ are monomorphic, epimorphic and isomorphic
  properties respectively, then the following formula is true\\
  $\forall(f\in Mor_{DB})(P_{mono}(f)\equiv P_{mono}(Tf)$ and $P_{epi}(f)\equiv
  P_{epi}(Tf)$  and  $P_{iso}(f)\equiv P_{iso}(Tf)$.
  \end{enumerate}
\end{theo}
\textbf{Proof:} It is easy to verify that $T$ is a 2-endofunctor and
to see that $T$ preserves properties of arrows: for example, if
$P_{mono}(f)$ is true for an arrow $f:A\longrightarrow B$, then
$\widetilde{f} = TA$ and $\widetilde{Tf}= T \widetilde{f} = T(TA) =
TA$, thus $P_{mono}(Tf)$ is true. Viceversa, if $P_{mono}(Tf)$ is
true then $\widetilde{Tf} =T \widetilde{f} = T(TA)$, i.e.,
$\widetilde{f} = TA$ and, consequently, $P_{mono}(f)$ is true.
\\$\square$\\
The endofunctor $T$ is a right and left adjoint to identity functor
$I_{DB}$, i.e., $T \simeq I_{DB}$. Thus we have  the equivalence
adjunction $< T, I_{DB}, \eta^C ,\eta >$ with the unit
$\eta^C:T\simeq I_{DB}$ (such that for any object $A$ the arrow
$\eta^C_A \triangleq \eta^C(A)\equiv is^{-1}_A:TA\longrightarrow
A$), and the counit $\eta:I_{DB} \simeq T$ (such that for any $A$
the arrow $\eta_A \triangleq \eta(A)\equiv is_A:A\longrightarrow
TA$)
are isomorphic arrows in $DB$ (by duality theorem it holds that $\eta^C = \eta^{inv}$).\\
The function $T^1:(A\longrightarrow
B)\longrightarrow(TA\longrightarrow TB)~$ is not a higher-order
function (arrows in $DB$ are not functions): thus, there is no
correspondent monad-comprehension for the monad $T$, which
invalidates the thesis ~\cite{Wald90} that "monads $\equiv~$
monad-comprehensions". It is only valid that "monad-comprehensions
$\Rightarrow~$ monads".\\
We have already seen that the views of a database can be seen as its
\emph{observable computations}: what we need, to obtain an
expressive power of computations in the category $DB$, are the
categorial computational properties, as known, based on monads:
\begin{propo} \label{prop:monad} The power-view closure 2-endofunctor $T =
(T^0,T^1):DB \longrightarrow DB~$ defines the monad $(T,\eta ,\mu)~$
and
  the comonad $(T,\eta^C ,\mu^C)~$ in DB, such that $\eta:I_{DB}\backsimeq
  T~$ and $\eta^C:T \backsimeq I_{DB}~$ are natural isomorphisms,
  while $\mu:TT \longrightarrow T~$ and $\mu^C:T \longrightarrow TT~$ are
  equal to the natural identity transformation $id_T:T \longrightarrow
  T~$ (because T = TT).
\end{propo}
\textbf{Proof:} It is easy to verify that all commutative diagrams
of the monad ($\mu_A \circ \mu_{TA} = \mu_A \circ T\mu_A~$, $\mu_A
\circ \eta_{TA} = id_{TA} = \mu_A \circ T\eta_A$) and the comonad
are diagrams composed by identity arrows. Notice that by duality we
obtain $\eta_{TA} = T\eta_A = \mu^{inv}_A$.
\\$\square$

\subsection{Duality}
The following duality theorem tells us that, for any commutative
diagram in $DB$, there is  the same commutative diagram composed by
equal objects and by inverted equivalent arrows as well. This
"bidirectional" mappings property of $DB$ is a consequence of the
fact that a composition of arrows is semantically based on the
set-intersection commutativity property for "information fluxes" of
its arrows. Thus \emph{any limit diagram} in $DB$ also has  its
\emph{"reversed" equivalent colimit diagram} with equal objects, and
\emph{any universal property} also has  its \emph{equivalent
couniversal property }in $DB$.
\begin{theo} \label{th:duality} there exists the controvariant functor
$\underline{S}= (\underline{S}^0,\underline{S}^1):DB \longrightarrow
DB$ such that
\begin{enumerate}
  \item $\underline{S}^0$ is an identity function on objects.
  \item for any arrow in $DB$, $f:A\longrightarrow B$ we have  $\underline{S}^1(f):B\longrightarrow
  A$, such that $\underline{S}^1(f) \triangleq f^{inv}$, where $f^{inv}$
  is an (equivalent) reversed morphism of $~f~~~$ (i.e., $\widetilde{f^{inv}}= \widetilde{f}$),\\
  $f^{inv}= is^{-1}_A \circ(Tf)^{inv}\circ is_B~$  with
\[
  \ (Tf)^{inv} ~\triangleq \bigcup_{ \partial_0(q_{TB_j})
   = \partial_1(q_{TB_j}) =\{v\} ~\& ~v \in~  \widetilde{f}} \{q_{TB_j}:TB
   \rightarrow TA \}
\]
  \item The category DB is equal to its dual category $DB^{OP}$.
\end{enumerate}
\end{theo}
\textbf{Proof:} We have, from the definition of reversed arrow,
that, $\widetilde{f^{inv}}=  \widetilde{is^{-1}_A}
\bigcap\widetilde{(Tf)^{inv}} \bigcap \widetilde{is_B}\\ = TA
\bigcap\widetilde{(Tf)^{inv}}\bigcap TB = TA\bigcap \widetilde{Tf}
\bigcap TB = TA\bigcap \widetilde{f} \bigcap TB = \widetilde{f}$.
The reversed arrow of any identity arrow is equal to it, and, also,
the compositional property for functor holds (the intersection
operator for "information fluxes" is commutative). Thus, the
controvariant functor is well defined.\\
It is convenient to represent this controvariant functor as a
covariant functor $S:DB^{OP}\\\longrightarrow DB$, or a covariant
functor $S^{OP}:DB\longrightarrow DB^{OP}$. It is easy to verify
that for compositions of these covariant functors hold, $SS^{OP}=
I_{DB}$ and $S^{OP}S= I_{DB^{OP}}$ w.r.t. the adjunction $< S,
S^{OP},\phi >:DB^{OP}\longrightarrow DB$, where $\phi~$ is a
bijection: for each pair of objects $A,B$ in $DB$ we have the
bijection of hom-sets, $\phi_{A,B}:DB(A,S(B))\simeq
DB^{OP}(S^{OP}(A),B)$, i.e., $\phi_{A,B}:DB(A,B)\simeq DB(B, A)$,
such that for any arrow $ f\in DB(A,B)~$ holds $\phi_{A,B}(f)
=S^1(f) = f^{inv}$. The unit and counit of this adjunction are the
identity natural transformations, $ \eta_{OP}:I_{DB}\longrightarrow
SS^{OP}$, $\epsilon_{OP}:S^{OP}S \longrightarrow  I_{DB^{OP}}~$
respectively, such that for any object $A$ they return by its
identity arrow. Thus, from this adjunction, we obtain that $DB$ is
isomorphic to its dual $DB^{OP}$; moreover they are \emph{equal}
because they have the same objects and the same arrows.
\\$\square$\\
 Let us
introduce the concepts for products and coproducts in $DB$ category.
\begin{definition} \label{def:coproduct}
The disjoint union  of any two instance-databases (objects) A and B,
denoted by $~A+B$, corresponds to two mutually isolated databases,
where two database management systems are completely disjoint, so
that it is impossible to compute the queries with
the relations from both databases.\\
The disjoint property for mappings is represented by facts that\\
$\partial_0(f+g)\triangleq \partial_0(f)+
\partial_0(g),~~~\partial_1(f+g)\triangleq \partial_1(f)+
\partial_1(g)$.
\end{definition}
Thus, for any database $A$, the \emph{replication} of this database
(over different DB servers) can be denoted by the coproduct object
$A + A$ in this category $DB$.
\begin{propo}\label{prop:disjiont}
For any two databases (objects) A and B we have that $T(A+B)= TA +
  TB$. Consequently $A + A$ is not isomorphic to $A$.
\end{propo}
\textbf{Proof:}  We have that $ T(A+B)= TA +
  TB$, directly from the fact that we are  able to define views
  only over relations in $A$ or, alternatively, over relations in $B$. Analogously $ ~~\widetilde{f+g} = \widetilde{f} +
  \widetilde{g}$, which is a closed object, that is, holds that $T(\widetilde{f+g}) = T(\widetilde{f} +
  \widetilde{g}) = T\widetilde{f} +
  T\widetilde{g} = \widetilde{f} +
  \widetilde{g} = \widetilde{f+g}$. \\
 From $T(A + A) = TA + TA \neq TA$ we obtain that $A + A$ is not isomorphic to $A$.
\\$\square$\\
Notice that for coproducts holds  that $~~ C + \perp^0 ~=~  \perp^0
+ C ~\simeq ~C$, and for any arrow $f$ in $DB$,
  $ ~~f + \perp^1 ~\approx~ \perp^1 + f ~\approx~
  f$, where $~\perp^1~$ is a banal empty morphism between objects, such that
  $~\partial_0(\perp^1)= \partial_1(\perp^1) = \perp^0$, with $\widetilde{\perp^1} =
  \perp^0$.\\
We are ready now to introduce the duality property between
coproducts and products in this $DB$ category:
\begin{propo}\label{prop:co-products} There exists an idempotent coproduct bifunctor
$+:DB\times DB\longrightarrow DB$ which is a disjoint union
operator for objects and arrows in DB.\\
The category DB is cocartesian with initial (zero) object $~\perp^0$
and for every pair of objects A,B it has a categorial coproduct
$A+B$ with monomorphisms (injections) $in_A:A
\hookrightarrow A+B~$ and $in_B:B \hookrightarrow A+B$.\\
By duality property we have that DB is also cartesian category with
a zero object $~\perp^0$. For each pair of objects A,B there exists
a categorial product $A\times B$ with epimorphisms (projections)
$p_A = in^{inv}_A:A\times A \twoheadrightarrow A~$ and $p_B =
in^{inv}_B:B\times B \twoheadrightarrow B$, where the product
bifunctor is equal to the coproduct bifunctor, i.e., $ \times~
\equiv~+$.
\end{propo}
\textbf{Proof:} 1. For any identity arrow $(id_A,id_B)$ in $DB\times
DB$, where $id_A ,~ id_b~$ are the identity arrows of $A$ and $B$
respectively, holds that $\widetilde{id_A+id_B} = \widetilde{id_A} +
\widetilde{id_B} = TA +TB =T(A+B) = \widetilde{id_{A+B}}$. Thus, $
+^1(id_A,id_B) = id_A+id_B = id_{A+B}$, is an identity arrow of the object $A+B$.\\
2. For any given $k:A\longrightarrow A_1$, $~k_1:A_1\longrightarrow
A_2$, $~l:B\longrightarrow B_1$, $~l_1:B_1\longrightarrow B_2$,
holds $\widetilde{+^1(k_1,l_1)\circ +^1(k,l)} =
\widetilde{+^1(k_1,l_1)}\bigcap \widetilde{+^1(k,l)}= \widetilde{k_1
\circ k + l_1 \circ
 l} = \widetilde{+^1(k_1 \circ k ,~ l_1 \circ l)}\\ = \widetilde{+^1((k_1, k) \circ(l_1,
 l))}$, thus $~+^1(k_1,l_1)\circ +^1(k,l) = +^1((k_1, k) \circ(l_1,
 l)).$\\
3. Let us demonstrate the coproduct property of this bifunctor: for
any two arrows $f:A\longrightarrow C$, $g:B\longrightarrow C$, there
exists a unique arrow $k:A+B\longrightarrow C$, such that $f =
k\circ in_A$, $g = k\circ in_B$, where $in_A:A \hookrightarrow A+B$,
$in_B:B \hookrightarrow A+B~$ are the injection (point to point)
monomorphisms ($\widetilde{in_A} = TA,~\widetilde{in_B}
= TB$). \\
It is easy to verify that for any two arrows $f:A\longrightarrow C$,
$g:B\longrightarrow C$,  there is exactly one arrow  $k = e_C
\circ(f+g):A+B \longrightarrow C$, where $e_C:C+C \twoheadrightarrow
C$ is an epimorphism (with $\widetilde{e_C} = TC$), such that
$\widetilde{k} = \widetilde{f} + \widetilde{g}$.
\\$\square$\\
The following proposition introduces the pullbacks (and pushouts, by
duality) for the category $DB$.
\begin{propo}\label{prop:pullback}  For any given pair of arrows with the same
codomain, $f:A\longrightarrow C$ and $g:B\longrightarrow C$, there
is a pullback with the fibred product $D = \widetilde{f}\bigcap
\widetilde{g}$ (product of A and B over C). By duality, for any pair
of arrows with the same domain there is a
pushout as well.\\
DB is a \textsl{complete} and \textsl{cocomplete} category.
\end{propo}
\textbf{Proof}: We define the commutative diagram $f \circ h_A =
h_B\circ g$, where $h_A:D \hookrightarrow A$ and  $h_B:D
\hookrightarrow B$ are monomorphisms defined by $h_A = is^{-1}_A
\circ in_{DA}$, $h_B = is^{-1}_B  \circ in_{DB}$, where
$is^{-1}_A:TA \longrightarrow A$, $~is^{-1}_B:TB \longrightarrow B$
are isomorphisms and $~in_{DA}:D\hookrightarrow TA$,
$~in_{DB}:D\hookrightarrow TB~$ are monomorphisms, such that
$\widetilde{h_A} = \widetilde{h_B}=
\widetilde{f}\bigcap \widetilde{g} = D $.\\
Let us show that for any pair of arrows $l_A:E\longrightarrow A$,
$l_B:E\longrightarrow B$, such that $f \circ l_A = l_B\circ g$ there
is a unique arrow $k:E\longrightarrow D$ such that a pullback
diagram
\begin{diagram}
E &          &              &              &               \\
  & \rdTo^k \rdTo(4,2)^{l_A}\rdTo(2,4)_{l_B}&   &  &        \\
  &          & D            & \rTo_{h_A}   &   A           \\
  &          &   \dTo_{h_B} &              &  \dTo_f        \\
  &          &  B           &  \rTo^g      &   C            \\
\end{diagram}
 commutes, i.e., (a) that $l_A = h_A \circ k~$ and $l_B
= h_B \circ k$. In fact, it must hold $\widetilde{k} \subseteq TD =
T(\widetilde{f}\bigcap \widetilde{g})= \widetilde{f}\bigcap
\widetilde{g} = \widetilde{h_A} = \widetilde{h_B}$. So, from the
commutativity (a), $\widetilde{l_A} = \widetilde{h_A} \bigcap
\widetilde{k} = \widetilde{k}~$ and $\widetilde{l_B} =
\widetilde{h_B} \bigcap \widetilde{k} = \widetilde{k}$. Thus , for
any other arrow $k_1:E \longrightarrow D$ that makes a commutativity
(a) must hold that $\widetilde{k_1} = \widetilde{l_A}
=\widetilde{l_B}$ and, consequently,
$\widetilde{k_1} = \widetilde{k}$, i.e., $k_1 = k$.\\
Consequently, $DB$ is a cartesian category with a terminal object
and pullbacks, thus it is complete (has all limits). By duality we
deduce that it is also cocomplete (has all colimits).
\\$\square$\\
In order to explain these concepts in another way, we can see the
limits and colimits as a left and a right adjunction for the
diagonal functor $\bigtriangleup:DB \longrightarrow DB^J$ for any
small index category (i.e., a diagram) $J$. For any colimit functor
$F:DB^J \longrightarrow DB$ we have a left adjunction to diagonal
functor $< F, \bigtriangleup, \eta_C, \varepsilon_C >:DB^J
\longrightarrow DB$, with the colimit object $F(D)$ for any object
(diagram) $D \in DB^J$ and the universal cone, a natural
transformation, $\eta_C:Id_{DB^J}\longrightarrow \bigtriangleup F$.
Then, by duality, the same functor $F$ is also a right adjoint to
the diagonal functor (adjunction, $<  \bigtriangleup, F, \eta,
\varepsilon >:DB \longrightarrow DB^J$), with the limit object
(equal to the colimit object above) $F(D)$ and the universal cone
(counit), a natural transformation, $\varepsilon:\bigtriangleup F
\longrightarrow Id_{DB^J}$, such that $\varepsilon = \eta^{inv}_C$
and $\eta = \varepsilon^{inv}_C$.\\
Let us see, for example, the coproducts ($F = +$) and products ($ F
= \times \equiv +$). In that case the diagram $D \in DB^J$ is just a
diagram of two arrows with the same codomain. We obtain for the
universal cone unit $\eta_C(<A,B>):<A,B> \longrightarrow <A+B,A+B>$
 one pair of coproduct inclusion-monomorphisms $\eta_C(<A,B>)=
<in_A , in_B>$, where $in_A:A \hookrightarrow A+B$, $in_B:B
\hookrightarrow A+B$. The universal cone counit of product
$\varepsilon (<A\times B, A\times B>):<A\times B, A\times
B>\longrightarrow <A,B>$ is  a pair of product
projection-epimorphisms $\varepsilon(<A\times B, A\times B>)=
<p_A,p_B>$, where $p_A:A\times B \twoheadrightarrow A$, $p_B:A\times
B \twoheadrightarrow B$, $A\times B = A+B$, $p_A = in^{inv}_A$, $p_B
= in^{inv}_B$, as represented in the following diagram:
\begin{diagram}
A            &          &   &                 &     A           \\
\dInto^{in_A}  & \rdTo^f  &   & \ruTo^{f^{inv}} &\uOnto_{p_A = in_A^{inv}}\\
A+B          & \rTo^k   & C & \rTo^{k^{inv}}  & A\times B  \\
\uInto^{in_B}  &  \ruTo_g &   & \rdTo_{g^{inv}} &  \dOnto_{p_B = in_B^{inv}}\\
 B           &          &   &                 &   B            \\
\end{diagram}
 \textbf{Example 3}: Let us verify that each object in $DB$ is a limit of
 some equalizer and a colimit of its dual coequalizer.  In fact,
 for any object $A$, a "structure map" $h:TA \longrightarrow A$ of
 a monadic T-algebra $<A, h>$ derived from a monad $(T, \eta, \mu)$ (where $h\circ \eta_A = id_A$,
  so that $h$ is an isomorphism $h = \eta^{inv}_A =\eta^C_A$, i.e.,
  $\widetilde{h} = TA = \widetilde{id_A}$)  we obtain the \emph{absolute
 coequalizer} (by Back's theorem, it is preserved by the
 endofunctor $T$, i.e., $T$ creates a coequalizer) with a colimit
 $A$, and, by duality, we obtain  the absolute equalizer with the
 limit $A$ as well.
\begin{diagram}
A&\lOnto^h &TA& \pile{\lTo^{Th}\\ \\\lTo_{\mu_A}}&T^2A& \pile{\lTo^{Th^{inv}}\\ \\\lTo_{\mu_A^{inv}}}&TA& \lInto^{h^{inv}}& A \\
\dTo^m & \ldTo_f& &  &    & & & \luTo_{f^{inv}} & \dTo_{m^{inv}}\\
B& & & & & & & & B \\
\end{diagram}
$\square$
\section{Equivalence relations for databases}
We can introduce a number of different equivalence relations for
instance-databases:
\begin{itemize}
  \item \emph{Identity} relation: Two  instance-databases (sets of relations)  $A$ and $B$ are
  identical when holds the set identity $A = B$.
  \item \emph{behavioral equivalence } relation: Two instance-databases   $A$ and $B$ are
  behaviorally equivalent when each view obtained from a database
  $A$ can also be  obtained from a database $B$ and viceversa.
  \item \emph{weak observational equivalence} relation: Two instance-databases   $A$ and $B$ are
  weakly equivalent when each "certain" view (without Skolem constants) obtained from a database
  $A$ can be also obtained from a database $B$ and viceversa.
\end{itemize}
It is also possible to define  other kinds of equivalences for
databases. In the rest of this chapter we will consider only the
second and third equivalences defined above.
\subsection{The (strong) behavioral equivalence for databases}
Let us  now consider the problem of how to define equivalent
(categorically isomorphic) objects (database instances) from a
\emph{behavioral point of view based on observations}: as we see,
each arrow (morphism) is composed by a number of "queries"
(view-maps), and each query may be seen as an \emph{observation}
over some database instance (object of $DB$). Thus, we can
characterize each object in $DB$ (a database instance) by its
behavior according to a given set of observations. Indeed, if one
object $A$ is considered as a black-box, the object $TA$ is only the
set of all observations on $A$. So, given two objects $A$ and $B$,
we are able to define the relation of equivalence between them based
on the notion of the bisimulation relation. If the observations
(resulting views of queries) of $A$ and $B$ are always equal,
independent of their particular internal structure, then
they look equivalent to an observer. \\
In fact, any database can be seen as a system with a number of
internal states that can be observed by using query operators (i.e,
programs without side-effects). Thus, databases $A$ and $B$ are
equivalent (bisimilar) if they have the same set of observations,
i.e. when $TA$ is equal to $TB$:
\begin{definition}\label{def:strong-eq}
The relation of (strong) behavioral equivalence $ '\approx'$ between
objects (databases) in $DB$ is defined by
\[ \ A \approx B~~iff~~TA = TB \]
  the equivalence relation for morphisms is given by,
\[f \approx g ~~~~iff~~~~ \widetilde{f} = \widetilde{g}\]
\end{definition}
 This relation of behavioral equivalence between objects corresponds to the notion of
isomorphism in the category $DB$ (see  Proposition \ref{prop:morphisms}).\\
This introduced  equivalence relation for arrows $\approx$, may be
given  by an (interpretation) function $B_T:Mor_{DB}\longrightarrow
Ob_{DB}~$ (see Definition \ref{semantix-morphism}), such that
$\approx$ is equal to the kernel of $B_T$, ($\approx~ =~ kerB_T$),
i.e., this is a fundamental concept for  categorial symmetry
~\cite{Majk98}:
\begin{definition}\label{def:symmetry} \textsc{Categorial symmetry}:\\
Let C be a category  with an \textsl{equivalence} relation $~\approx
~\subseteq Mor_C \times Mor_C$
 for its arrows (equivalence
relation for objects is the isomorphism $~\backsimeq ~\subseteq Ob_C
\times Ob_C$) such that there exists a bijection between equivalence
classes of $\approx$ and $\backsimeq$, so that it is possible to
define a skeletal category $|C|$ whose objects are defined by the
imagine of a function $B_T:Mor_C \longrightarrow Ob_C$ with the
kernel  $~kerB_T = ~ \thickapprox$, and to define an associative
composition operator for objects $*$,  for any fitted pair
$g\circ f$ of arrows, by $~~ B_T(g) * B_T(f)= B_T(g\circ f)$.\\
For any arrow in C, $f:A\longrightarrow B$, the object $B_T(f)$ in
C, denoted by $~\widetilde{f}$, is denominated as a
\textsl{conceptualized} object.
 \end{definition}
 \textbf{Remark:} This symmetry property allows us to consider all the properties
 of an arrow (up to the equivalence) as properties of objects and their
composition as well. Notice that any two arrows are \textsl{equal}
if and only if they are equivalent and have the same source and the
target objects.\\
We have that in symmetric categories holds that $f \approx g$ iff
$\widetilde{f} \simeq \widetilde{g}$.\\
 Let
us introduce, for a category $C$ and its arrow category $C\downarrow
C$, an encapsulation operator $J:Mor_C\longrightarrow
Ob_{C\downarrow C}$, that is,
 a one-to-one function such that for any arrow $f:A\longrightarrow B$, $J(f) = <A,B,f>$ is its correspondent object in $C\downarrow C$,
 with  its inverse $~\psi~ $ such that $\psi(<A,B,f>) = f$.\\ We
 denote by  $F_{st},S_{nd}:(C \downarrow
C)\longrightarrow C ~$  the first and the second comma functorial
projections (for any functor $F:C \rightarrow D$ between categories
$C$ and $D$, we denote by $F^0$ and $F^1$ its object and arrow
component), such that for any arrow $(k_1; k_2):<A,B,f>\rightarrow
<A',B',g>$ in $C\downarrow C$ (such that $k_2 \circ f = g \circ k_1$
in $C$), we have that $F_{st}^0(<A,B,f>) = A, F_{st}^1(k_1; k_2) =
k_1$ and
$S_{nd}^0(<A,B,f>) = B, S_{nd}^1(k_1; k_2) = k_2$.\\
We denote by $\blacktriangle :C\longrightarrow (C\downarrow C)~$ the
 diagonal functor, such that for any object $A$ in a category $C$, $\blacktriangle^0(A) =
 <A,A,id_A>$.\\
 An important subset of symmetric categories are Conceptually
 Closed and Extended symmetric categories, as follows:
\begin{definition}\label{def:symCat} \textsl{Conceptually closed} category is a symmetric category C with a functor
$T_e = (T_e^0,T_e^1): (C\downarrow C)\longrightarrow C$
such that  $T_e^0 = B_T \psi$, i.e., $B_T = T_e^0 J$,
 with a natural isomorphism $~\varphi:T_e \circ \blacktriangle~
\backsimeq~I_{C}$, where $I_C$ is an identity functor for $C$.\\
 C is an \textsl{extended symmetric} category if  holds also $~~~\tau^{-1}\bullet \tau = \psi$, for  vertical
 composition of natural transformations $\tau: F_{st} \longrightarrow
T_e ~$ and $ \tau^{-1}:T_e \longrightarrow S_{nd} $.
\end{definition}
Remark: it is easy to verify that in conceptually closed categories,
it holds that   any arrow $f$ is equivalent to an \emph{identity} arrow, that is, $f \approx id_{\widetilde{f}}$.\\
It is easy  to verify also that in extended symmetric categories the following holds:\\
$~~ \tau = (T_e^1 (\tau_I F_{st}^{0} ;\psi ))\bullet
 (\varphi^{-1}F_{st}^{0}) ,~~ \tau^{-1} = (\varphi^{-1} S_{nd}^{0}) \bullet (T_e^1 ( \psi ;\tau_I S_{nd}^{0}
 ))$,\\ where $\tau_I : I_{C}\longrightarrow
 I_{C}~$ is an identity natural transformation (for any object $A$ in $C$, $\tau_I(A)
 =id_A$).\\
 \textbf{Example 4:} The $Set$ is an extended symmetric
category: given any function $~f:A\longrightarrow B$ , the
conceptualized object  of this function is the graph of this
function (which is a set), $\widetilde{f}= B_T(f) = \{(x,f(x))~|~x
\in A\}$.\\
The equivalence $\approx$ on morphisms (arrows) is defined by: two
arrows $f$ and $g$ are equivalent, $f \approx g$, iff they have the
same graph.\\ The composition of objects $*$ is defined as
associative composition of binary relations (graphs), $B_T(g \circ
f) = \{(x,(g\circ f)(x))~|~x \in A\} =  \{(y,g(y))~|~y \in B\} \circ
\{(x,f(x))~|~x \in A\} = B_T(g) * B_T(f)$.\\
$Set$ is also conceptually closed by the functor $T_e$, such that
for any object $J(f) = <A,B,f>$, $T_e^0(J(f)) = B_T(f) =
\{(x,f(x))~|~x \in A\}$, and for any arrow
$(k_1;k_2):J(f)\rightarrow J(g)$, the component $T_e^1$ is defined
by: \\for any $(x,f(x)) \in T_e^0(J(f)), ~~T_e^1(k_1; k_2)(x,f(x)) =
(k_1(x), k_2(f(x)))$.\\ It is easy to verify the compositional
property for $T_e^1$, and that $T_e^1(id_A; id_B) =
id_{T_e^0(J(f))}$. For example, $Set$ is also an extended symmetric
category, such that for any object $J(f) = <A,B,f>$ in $Set
\downarrow Set$, we have that $\tau(J(f)):A \twoheadrightarrow
B_T(f)$ is an epimorphism, such that for any $x \in A$,
$~~\tau(J(f))(x) = (x, f(x))$, while $\tau^{-1}(J(f)):B_T(f)
\hookrightarrow B$ is a monomorphism such that for any $(x, f(x))
\in B_T(f), \\~~\tau^{-1}(J(f))(x, f(x)) = f(x)$.\\
Thus, each arrow in $Set$ is a composition of an epimorphism and a
monomorphism.
\\$\square$\\
 Now we are ready to present a
formal definition for the $DB$ category:
\begin{theo}\label{th:symmetry} The category DB is an extended symmetric category, closed by
the functor $T_e = (T_e^0,T_e^1): (C\downarrow C)\longrightarrow C$,
where $T_e^0 = B_T \psi~$ is the object component of this functor
such that for any arrow $f$ in DB, $T_e^0(J(f)) = \widetilde{f}$,
while its arrow component $~T_e^1$ is defined as follows: for any
arrow $(h_1;h_2):J(f)\longrightarrow J(g)~$ in $DB\downarrow
  DB$, such that $g\circ h_1 = h_2\circ f~$ in DB, holds
\[
\ T_e^1(h_1;h_2) =  \bigcup_{ \partial_0(q_{\widetilde{f}_i})
=\partial_1(q_{\widetilde{f}_i}) =\{v\} ~\& ~v \in ~ \widetilde{h_2
\circ f}} \{q_{\widetilde{f}_i}\}
\]
The associative composition operator for objects $*$, defined for
any fitted pair $g\circ f$ of arrows, is the set intersection
operator $\bigcap$. \\Thus, $~~ B_T(g)
* B_T(f) = \widetilde{g}\bigcap \widetilde{f} = \widetilde{g\circ f} = B_T(g\circ
f)$.
\end{theo}
\textbf{Proof:}  Each object $A$ has its identity (point-to-point)
morphism
$ id_A = \\ \bigcup_{ \partial_0(q_{A_i}) =\partial_1(q_{A_i})
=\{v\} ~\& ~v \in A} \{q_{A_i}\}$
and holds the associativity  $\widetilde{h\circ (g\circ f)}=
\widetilde{h} ~\bigcap ~\widetilde{(g\circ f)}\\ = \widetilde{h}
~\bigcap ~\widetilde{g }~ \bigcap~ \widetilde{f}= \widetilde{(h\circ
g)}~\bigcap ~ \widetilde{f} = \widetilde{(h\circ g)\circ f}$. They
have the same source and target object, thus $h\circ(g\circ f)=
(h\circ g)\circ f$. Thus, $DB$ is a category. It is easy to verify
that also $T_e~$ is a well defined functor. In fact, for any
identity arrow $(id_A;id_B):J(f) \longrightarrow J(f)$ it holds that
$T_e^1(id_A;id_B) =  \bigcup_{
\partial_0(q_{\widetilde{f}_i}) =\partial_1(q_{\widetilde{f}_i})
=\{v\} ~\& ~v \in ~ \widetilde{id_B \circ f}}
\{q_{\widetilde{f}_i}\} = id_{\widetilde{f}} $ is the identity arrow
of $\widetilde{f}$. For any two arrows
$(h_1;h_2):J(f)\longrightarrow J(g)$, $(l_1;l_2):J(g)\longrightarrow
J(k)$, it holds that $\overline{T_e^1(h_1;h_2)\circ T_e^1(l_1;l_2)}
= \widetilde{T_e^1(h_1;h_2)}\bigcap \widetilde{T_e^1(l_1;l_2)} =
T(\widetilde{l_2\circ g})\bigcap T(\widetilde{h_2\circ f})=
\widetilde{l_2}\bigcap \widetilde{g}\bigcap \widetilde{h_2}\bigcap
\widetilde{f} = (by~l_2\circ f = g\circ h_1)=\widetilde{l_2}\bigcap
\widetilde{g}\bigcap \widetilde{h_1}\bigcap \widetilde{h_2}\bigcap
\widetilde{f} = (by~l_2\circ f = g\circ h_1)= \widetilde{l_2}\bigcap
 \widetilde{h_2}\bigcap \widetilde{f} = \widetilde{l_2\circ h_2 \circ f}= T_e^1(l_1\circ h_1;l_2\circ h_2)$,
 finally, $T_e^1(h_1;h_2)\circ T_e^1(l_1;l_2)= T_e^1(l_1\circ h_1;l_2\circ
 h_2)$. For any identity arrow, it holds that $id_A$, $T_e^0J(id_A) = \widetilde{id_A}
 = TA \simeq A$ as well, thus,  an isomorphism $~\varphi:T_e \circ \blacktriangle~
\backsimeq~I_{DB}$ is valid.
\\$\square$\\
\textbf{Remark:} It is easy to verify (from $~~\tau^{-1}\bullet \tau
= \psi$) that for any given morphism $~f:A\longrightarrow B~$ in
$DB$, the arrow $f_{ep} = \tau (J(f)):A \twoheadrightarrow
\widetilde{f}~$ is an epimorphism, and the arrow $f_{in} =
\tau^{-1}(J(f)):\widetilde{f} \hookrightarrow B~$ is a monomorphism,
so that \emph{any morphism f in DB is a composition of an
epimorphism and  monomorphism} $ f = f_{in} \circ f_{ep}$, with the
intermediate
object equal to its "information flux" $\widetilde{f}$, and with $f \approx f_{in} \approx f_{ep}$.\\
Let us prove that the equivalence relations on objects and morphisms
are based on the "inclusion" Partial Order (PO) relations, which
define the $DB$ as a 2-category:
\begin{propo}\label{prop:POCat}  The subcategory $DB_I \subseteq DB$ , with $Ob_{DB_I}
= Ob_{DB}$ and with  monomorphic arrows only, is a Partial Order
category with the PO relation of "inclusion" $A \preceq B$ defined
by a monomorphism $f:A\hookrightarrow B$.  The "inclusion" PO
relations for objects and arrows are defined as follows:
\[
 A \preceq B ~~~~ iff ~~~~ TA \subseteq TB
\]
\[
 f \preceq g ~~ ~~iff ~~~~ \widetilde{f } \preceq
 \widetilde{g}~~~~(i.e.,\widetilde{f } \subseteq
 \widetilde{g}~~)
\]
      they determine two observation equivalences, i.e.,
\[
 A \backsimeq B ~~(i.e.,~~ A \thickapprox B)~~~~~~ iff
 ~~ ~~A\preceq B~~and ~~B \preceq A
\]
\[
 f \thickapprox g ~~~~ iff ~~~~ f  \preceq g ~~and ~~ g \preceq f
 ~
\]
The power-view endofunctor $T:DB\longrightarrow DB$ is a
2-endofunctor and a closure operator for this PO relation: any
object A such that $A = TA$ will be called a "closed object".\\
 $DB$ is a 2-category, 1-cells are its ordinary morphisms, while
2-cells (denoted by $\sqrt{\_}~$) are the arrows between ordinary
morphisms: for any two morphisms $~f,g:A\longrightarrow B$ , such
that $f \preceq g~$, a 2-cell arrow is the "inclusion"
$\sqrt{\alpha}:f\frac{\preceq}{\longrightarrow}g $. Such a 2-cell
arrow is represented by an ordinary arrow in $DB$,
$\alpha:\widetilde{f} \hookrightarrow \widetilde{g}$, where $\alpha
= T_e^1(id_A; id_B)$.
\end{propo}
\textbf{Proof:} The relation $A \preceq B$ is well defined:
 any monomorphism $f:A\hookrightarrow B$  is a unique monomorphism
 (for any other monic arrow $g:A\hookrightarrow B$ must hold $\widetilde{g} =
 TA = \widetilde{f}$, thus $g = f$). Consequently,  between any two
 given objects in $DB_I$ there can exist at maximum one arrow, so this is a PO
 category. The "inclusion" $A \preceq B$ is not a simple set
 inclusion $ ~\subseteq~ $
 between elements of $A$ and elements of $B$ (this is the case only for
 closed objects and, generally, $A \subseteq B$ implies $A \preceq B$, but not viceversa). The following properties are valid:
\begin{enumerate}
  \item $A \preceq B$ implies $ TA \preceq TB ~~( i.e., TA \subseteq
  TB)$, from the definition of $\preceq$, if all elements of $A$ can
  define only one part of $B$, then the set of views of $A$ is a
  subset of the set of views of $B$:  $T$ is a monotonic operator.
  \item $A \preceq TA$, i.e., each element of $A$ is also a view
  of $A$.
  \item $TA =TTA$, as explained at the beginning of this
  paper.
\end{enumerate}
Thus, $T$ is a closure operator, and an object $A$, such that  $A =
TA$ is a closed object. The rest of the proof comes directly from
Proposition \ref{prop:morphisms} and the definitions.
 Let us verify that  the arrow component of this
endofunctor is a closure operator as well:
\begin{enumerate}
  \item $f \preceq g$ implies $ Tf \preceq Tg ~~$( i.e.,  from $f \preceq g$
  holds that $ \widetilde{f} \subseteq \widetilde{g}$, thus $ T\widetilde{f} \subseteq
  T\widetilde{g}$, i.e. $ \widetilde{Tf} \subseteq  \widetilde{Tg}$)
  \item $f \preceq Tf$, from $\widetilde{f} \preceq \widetilde{Tf}~~~$
  \item $Tf =TTf$, in fact $ \widetilde{Tf} = T\widetilde{f} =
  TT\widetilde{f} = T\widetilde{Tf} = \widetilde{TTf}$
\end{enumerate}
Notice that for each arrow $f$ it holds (by  closure property  of
$T$ that $f \approx Tf$, i.e.,
that $\widetilde{f }= T\widetilde{f} = \widetilde{Tf}$.\\
 It is easy to verify that $DB$ is a 2-category with 0-cells (its
 objects), 1-cells (its ordinary morphisms (mappings)) and
 2-cells (arrows ("inclusions") between mappings). The horizontal
 and  vertical composition of 2-cells is just the composition of PO
 relations $\preceq$ :  given $~f,g,h:A\longrightarrow B$ with 2-cells
 $\sqrt{\alpha}:f\frac{\preceq}{\longrightarrow}g$,$~\sqrt{\beta}:g\frac{\preceq}{\longrightarrow}h$,
  then their vertical composition is $\sqrt{\gamma} = \sqrt{\beta}
\circ  \sqrt{\alpha}:f\frac{\preceq}{\longrightarrow}h~$; given
$~f,g:A\longrightarrow
 B~$ and $~h,l:B\longrightarrow C$, with 2-cells $~\sqrt{\alpha}:f\frac{\preceq}{\longrightarrow}~g$,
 $\sqrt{\beta}:h\frac{\preceq}{\longrightarrow}~l$, then, for a
given
  composition functor $\bullet :DB(A,B)\times DB(B,C)\longrightarrow
 DB(A,C)$, their horizontal composition is $\sqrt{\gamma} = \sqrt{\beta} \bullet
 \sqrt{\alpha}:h\circ f~\frac{\preceq}{\longrightarrow}~l\circ g$.
\\$\square$\\
\textbf{Example 5:} Equivalent morphisms: for any view-map
$q_{A_i}:A\longrightarrow TA$
the equivalence with another view-mapping $q_{B_j}:B\longrightarrow
TB $ is obtained
when they produce the same
view.\\
Let us  now see  that each 2-cell may be represented by an
equivalent ordinary morphism (1-cell) (from $~f \preceq g ~~ ~~iff
~~~~ \widetilde{f } \preceq\widetilde{g}$), and moreovr, that we are
able to treat the \emph{mappings between mappings} directly as
morphisms of  the $DB$ category.\\
The categorial symmetry operator $~T_e^0J: Mor_{DB}\longrightarrow
Ob_{DB}~~$ for any mapping (morphism) $f$ in $DB$ produces its
"information flux" object $\widetilde{f}$ (i.e., the
"conceptualized" database of this mapping). Consequently, we can
define a "mapping between mappings"  (which are 2-cells
("inclusions")) and also all higher n-cells ~\cite{BaDo98} by their
direct transposition into a 1-cell morphism, but  we are able to
make more complex morphisms between mappings as well.
\\$\square$\\
\textbf{Example 6:} Let us consider the two ordinary (1-cells)
 morphisms in $DB$,  $~f:A\longrightarrow B$, $~g:C\longrightarrow D~$
 such that $f\preceq g $.
 We want to show that its 1-cells correspondent monomorphism $\alpha:\widetilde{f}\hookrightarrow \widetilde{g}
 ~~$ is a  result of the symmetric closure functor $T_e$.
 Let us
 prove that for two arrows, $h_A = is_C\circ in_C\circ \tau(J(f))~$ and $h_B = is_D\circ in_D\circ e_B \circ is_B
 ~$  (where   $in_C:\widetilde{f} \hookrightarrow TC $ is a monomorphism (well defined,
  because $f\preceq g$ implies
 $\widetilde{f }\subseteq \widetilde{g}\subseteq TC$), $is_C:TC \longrightarrow C $ is an isomorphism,
 $is_B:B \longrightarrow TB $ is an isomorphism, $e_B:TB \twoheadrightarrow \widetilde{f}$
 is an epimorphism, $in_D:\widetilde{f} \hookrightarrow TD $ is a monomorphism
 ($\widetilde{f }\subseteq \widetilde{g}\subseteq TD$),
  $is_D:TD \longrightarrow D $ is an isomorphism) holds that $g\circ h_A = h_B \circ f~$: we have that
   $ \widetilde{h_A} = \widetilde{is_C} \bigcap \widetilde{in_C} \bigcap \widetilde{\tau(J(f))}
  = TC \bigcap T\widetilde{f} \bigcap  \widetilde{f} = \widetilde{f}$,
  ( because $TC \supseteq \widetilde{g} \supseteq \widetilde{f}~
   ~and ~T\widetilde{f} = \widetilde{f}$), and analogously $ \widetilde{h_B}
   = T\widetilde{f} = \widetilde{f}$. Thus, $\widetilde{g\circ h_A} = \widetilde{h_B \circ f} =
   \widetilde{f}$, and finally $g\circ h_A = h_B \circ f$.\\
   Thus,  there exists the arrow $(h_A;h_B):J(f)\longrightarrow J(g) $ in $DB\downarrow
   DB$. Let us prove that also $T_e^1(h_A;h_B)$ is a monomorphism as well,
   and that it holds that $\alpha = T_e^1(h_A;h_B):\widetilde{f}\hookrightarrow \widetilde{g}
   $: in fact, by definition,
\[
\ T_e^1(h_A;h_B) =  \bigcup_{ \partial_0(q_{\widetilde{f}_i})
=\partial_1(q_{\widetilde{f}_i}) =\{v\} ~\& ~v \in ~ \widetilde{h_B
\circ f}} \{q_{\widetilde{f}_i}\} = \bigcup_{
\partial_0(q_{\widetilde{f}_i}) =\partial_1(q_{\widetilde{f}_i}) =\{v\} ~\& ~v \in
~ \widetilde{f}} \{q_{\widetilde{f}_i}\}
\]
 because $ \widetilde{h_B \circ f} =\widetilde{f} $.
  Thus, $ \widetilde{T_e^1(h_A;h_B)} = T\widetilde{f} = \widetilde{f} $ and,  consequently, $ T_e^1(h_A;h_B)$ is
  a  monomorphism.\\
 In the particular case  when $A = C$ and $B = D$ we obtain for
 the 2-cells arrow $\sqrt{\alpha}:f\frac{\preceq}{\longrightarrow}g
 ~~$ represented by the 1-cell arrow $\alpha = T_e^1(id_A;id_B): \widetilde{f} \hookrightarrow \widetilde{g}$.
 \\$\square$
 %
 \subsection{Weak  observational equivalence for databases }
A database instance can also have relations with tuples containing
\emph{Skolem constants} as well (for example, the minimal Herbrand
models for Global (virtual) schema of some Data
integration system ~\cite{Lenz02,CCGL02,FKMP03}).\\
In what follows we consider a recursively enumerable set of all
Skolem constants as marked (labeled) nulls $SK = \{ \omega_0,
\omega_1,...\}$, disjoint from a domain set \textbf{dom} of all
values for databases, and we introduce a unary predicate $Val(\_)$,
such that $Val(t)$ is true iff $t \in \textbf{dom}$
(so, $Val(\omega_i)$ is false for any $\omega_i \in SK$).\\
Thus, we can define a new weak power-view operator for databases as
follows:
\begin{definition} \label{def:weakendofunctor}
Weak power-view operator $T_w:Ob_{DB}\longrightarrow Ob_{DB}$ is
defined as follows: for any database $A$ in $DB$ category it
holds that:\\
$T_w(A) \triangleq \{~v~|~v \in T(A)~and~\forall_{1\leq k \leq
|v|}\forall(t\in \pi_k(v))Val(t)\}$\\
 where $|v|$ is the number of attributes of the view $v$, and
 $\pi_k$ is a k-th projection operator on relations.\\
 We define a partial order relation $\preceq_w$ for databases:\\
 $~~~~~~A \preceq_w B ~~$ iff $~~T_w(A) \subseteq T_w(B)$\\
 and we define a weak observational equivalence relation
 $\approx_w$ for databases:\\
$~~~~~~A \approx_w B ~~$ iff $~~A\preceq_w B~ and ~B\preceq_w A$.
\end{definition}
The following properties hold for the weak partial order
$\preceq_w$, w.r.t. the partial order $\preceq$ (we denote $'A\prec
B'$ iff $A\preceq B$ and not $A \simeq B$):
\begin{propo} \label{prop:weak}
Let $A$ and $B$ be any two databases (objects in $DB$ category),
then:
\begin{enumerate}
 \item $T_w(A)\simeq A$, if $A$ is a database without Skolem
  constants\\
$T_w(A)\prec A$, otherwise
\item $A \prec B$ implies $A \preceq_w B$
\item $ A \simeq B$ implies $A \approx_w B$
    \item $T_w(T_w(A)) = T(T_w(A))= T_w(T(A)) = T_w(A) \subseteq
  T(A)$\\
 thus, each object $D = T_w(A)$ is a closed object (i.e., $D =
  T(D)$) such that $D\approx_w A$
  \item $T_w$ is a closure operator w.r.t. the "weak inclusion"
  relation $\preceq_w$
 \end{enumerate}
\end{propo}
\textbf{Proof:} 1. From $T_w(A)\subseteq T(A)$ ($T_w(A)= T(A)$ only
if $A$ is
without Skolem constants). \\
2. If $A \prec B$ then $T(A)\subset T(B)$, thus
$T_w(T(A))\subseteq T_w(T(B))$, i.e., $A \preceq_w B$.\\
 3. Directly from (4) and the fact that $A\simeq B$ iff $A \preceq
 B$ and $B \preceq A$.\\
 4. It holds from definition of the operator $T$ and $T_w$:
  $T_w(T_w(A)) = T(T_w(A))$ because $T_w(A)$ is the set of views of $A$
 without Skolem constants and from (1). $T_w(T(A)) = \{~v~|~v \in TT(A)~and~\forall_{1\leq k \leq
|v|}\forall(t\in \pi_k(v))Val(t)\} = \{~v~|~v \in
T(A)~and~\forall_{1\leq k \leq |v|}\forall(t\in \pi_k(v))Val(t)\} =
T_w(A)$, from $T =TT$. Let us show that $T_w(T_w(A)) = T_w(A)$. For
every view $v \in T_w(T_w(A))$,  from $T_w(T_w(A)) = T(T_w(A))
\subseteq TA$,
holds that $v \in TA$ and from the fact that $v$ is without Skolem constants it follows that $v\in T_w(A)$. The converse is obvious.\\
5. We have that $A \preceq_w T_w(A)$, $A \preceq_w B$ implies
$T_w(A)\preceq_w T_w(B)$, and $T_w(T_w(A)) = T_w(A)$. Thus, $T_w$ is
a closure operator.\\$\square$\\
Notice that from point 4, the partial order $"\preceq"$ is a
stronger discriminator for databases than the weak partial order
$"\preceq_w"$, i.e., we can have two non isomorphic objects $ A
\prec B$ that are weakly equivalent, $A\approx_w B$ (for example
when $A = T_w(B)$ and $B$ is a database with Skolem constants). Let
us extend the notion of the type operator $T$ into the notion of the
endofunctor of $DB$ category:
\begin{theo} \label{th:weakendofunctor}There exists the weak power-view endofunctor $T_w =
(T_w^0,T_w^1): DB \longrightarrow DB$, such that
\begin{enumerate}
  \item for any object A, the object component $T_w^0$ is
  equal to the type operator $T_w$.
  \item for any morphism $~f:A\longrightarrow B$, the arrow
  component $T_w^1$ is  defined by
  \[
  \ T_w(f) \triangleq T_w^1(f) = inc_B^{inv} \circ T^1(f)\circ inc_A
  \]
  where $inc_A:T_w(A)\hookrightarrow T(A)$ is a monomorphism (set
  inclusion) and $inc_B^{inv}:T(B)\twoheadrightarrow T_w(B)$ is an
  epimorphism (reversed monomorphism $inc_B$).
  \item Endofunctor $T_w$ preserves the properties of arrows, i.e., if a
  morphism $f$ has a property P (monic, epic, isomorphic), then
  also $T_w(f)$ has the same property: let $P_{mono} , P_{epi}$
  and $P_{iso}$ are monomorphic, epimorphic and isomorphic
  properties respectively, then the following formula is true\\
  $\forall(f\in Mor_{DB})(P_{mono}(f)\equiv P_{mono}(T_w f) \wedge P_{epi}(f)\equiv P_{epi}(T_w f) \wedge
  P_{iso}(f)\equiv P_{iso}(T_w f )$.
  \item There exist the natural transformations,
  $\xi:T_w\longrightarrow T$ (natural monomorphism), and $\xi^{-1}:T\longrightarrow T_w$ (natural
  epimorphism), such that for any object $A$, $\xi(A)= inc_A$ is a monomorphism and $\xi^{-1}(A)=
  inc_A^{inv}$ is an epimorphism such that $\xi(A)\approx
  \xi^{-1}(A)$.
  \end{enumerate}
\end{theo}
\textbf{Proof:} It is easy to verify that for any two arrows $f:A
\longrightarrow B$, $g:B\longrightarrow C$, it holds that
$\widetilde{T_w(g\circ f)}\subseteq T(T_w(B)= \widetilde{inc_B\circ
inc_B^{-1}}$, thus $T_w(g\circ f) = inc_C^{-1}\circ T^1(g\circ
f)\circ inc_A = inc_C^{-1}\circ T^1(g) \circ T^1(f)\circ inc_A =
inc_C^{-1}\circ T^1(g)\circ inc_B\circ inc_B^{-1} \circ T^1(f)\circ
inc_A = T_w(g) \circ T_w(f)$. Thus, it is an endofunctor. The rest
is easy to verify.
\\$\square$\\
Like the monad $(T, \eta , \mu)~$  and comonad $(T,\eta^C ,\mu^C)~$
of the endofunctor $T$, we can define such structures for the weak
endofunctor $T_w$ as well:
\begin{propo} \label{prop:weakmonad} The weak power-view endofunctor $T_w =
(T_w^0,T_w^1):DB \longrightarrow DB~$ defines the monad $(T_w,\eta_w
,\mu_w)~$ and
  the comonad $(T_w,\eta^C_w ,\mu^C_w)~$ in $DB$, such that $\eta_w = \xi^{-1} \bullet \eta:I_{DB}
  \longrightarrow T_w$ is a natural epimorphism
   and $\eta^C_w = \eta^C\bullet \xi:T_w \longrightarrow I_{DB}~$ is a natural monomorphisms
   ($~'\bullet'$ is a vertical composition for natural transformations),
  while $\mu_w :T_wT_w \longrightarrow T_w~$ and $\mu^C_w :T_w \longrightarrow T_wT_w~$ are
  equal to the natural identity transformation $id_{T_w} :T_w \longrightarrow
  T_w~$(because $T_w = T_wT_w$).
\end{propo}
\textbf{Proof:} It is easy to verify that all commutative diagrams
of the monad and the comonad are diagrams composed by identity
arrows.
\\$\square$

\section{Categorial Semantics for Data Integration/Exchange }
Data exchange~\cite{FKMP03} is a problem of taking data structured
under a source \emph{schema} and creating an instance of a target
\emph{schema} that reflects the source data as accurately as
possible. Data integration ~\cite{Lenz02} instead is a problem of
combining data residing at different sources, and providing the user
with a unified global \emph{schema} of this data. Thus, in this
framework the concepts are defined in a more abstract way than in
the instance database framework represented in the "computation"
$DB$ category. Consequently, we require an interpretation mapping
from the scheme into the instance level, which will be given
categorially by functors.
\subsection{Data Integration/Exchange Framework }
We formalize a \emph{data integration system} $\I$ in terms of a
triple $\tup{\G,\S,\M}$, where
\begin{itemize}
\item $\G = (\G_T, \Sigma_T)$ is the \emph{target schema}, expanded by the \emph{new unary predicate} $Val(\_)$ such that
 $Val(c)$ is true if $c \in \textbf{dom}$, expressed in a language $\L_{\G}$
  over an alphabet $\A_\G$, where $\G_T$ is the schema and $\Sigma_T~$ are its integrity constraints.
  The alphabet comprises a symbol for each element
  of $\G$ (i.e., relation if $\G$ is relational, class if $\G$ is
  object-oriented, etc.).
\item $\S$ is the \emph{source schema}, expressed in a language
  $\L_{\S}$ over an alphabet $\A_\S$. The alphabet $\A_\S$ includes a
  symbol for each element of the sources. While the source
  integrity constraints may play an important role in deriving
  dependencies in $\M$,
  they do not play any direct
  role in the data integration/exchange framework and we may ignore
  them.
\item $\M$ is the \emph{mapping} between $\G$ and $\S$, constituted by a set of
  \emph{assertions} of the forms
  \begin{center}
     $(1)~~~\Map{q_{\S}}{q_{\G}}$,
     $~~\Map{q_{\G}}{q_{\S}}$
   \end{center}
   where $q_{\S}$ and $q_{\G}$ are two queries of the same arity,
    over the source schema $\S$ and over the target schema $\G$ respectively.
   Queries $q_{\S}$ are expressed in a query language $\L_{\M,\S}$ over the
   alphabet $\A_\S$, and queries $q_{\G}$ are expressed in a query language
   $\L_{\M,\G}$ over the alphabet $\A_\G$.  Intuitively, an assertion
   $\Map{q_{\S}}{q_{\G}}$ specifies that the concept represented by the query
   $q_{\S}$ over the sources corresponds to the concept in the target schema
   represented by the query $q_{\G}$ (similarly for an assertion
   of type $\Map{q_{\G}}{q_{\S}}$).
\item Queries $q_C(\textbf{x})$, where $\textbf{x} = {x_1,..,x_k}$ is a non empty set of variables,
over the global schema are \emph{conjunctive queries}. We will use,
for every original query $q_C(\textbf{x})$, only a \emph{lifted
query} over the global schema, denoted by $q$, such that $q :=
q_C(\textbf{x})\wedge Val(x_1) \wedge ...\wedge Val(x_k)$.
\end{itemize}
In order to define the semantics of a data integration system, we
start from the data at the sources, and specify which are the data
that satisfy the global schema. A \emph{source database} $\D$ for
$\I=\tup{\G,\S,\M}$ is constituted by one relation $\ext{r}{\D}$ for
each source $r$ in $\S$ (sources that are not relational may be
suitably presented in the relational form by wrapper's programs). We
call \emph{global
  database} for $\I$, or simply {\em database} for $\I$, any database for $\G$.
A database $\B$ for $\I$ is said to be \emph{legal} with respect to
$\D$ if:
\begin{itemize}
\item $\B$ satisfies the integrity constraints of $\G$;
\item $\B$ satisfies $\M$ with respect to $\D$.
\item We restrict our attention to sound views only, which are typically considered the
most natural ones in a data integration
setting~\cite{Lenz02,Hale01}.
\end{itemize}
In order to obtain an answer to a lifted query $q$ from a data
integration system, a tuple of constants is considered an answer to
this query only if it is a \emph{certain}  answer, i.e., it
satisfies
the query in \emph{every legal} global database.\\
We may try to infer all the legal databases for $\I$  and compute
the tuples that satisfy the lifted query
  $q$ in all such legal databases. However, the difficulty here is that, in
  general, there is an infinite number of legal databases. Fortunately
  we can define another \emph{universal(canonical)} database $\can(\I,\D)$,  that has the interesting property
  of faithfully representing all legal databases. The construction of the canonical database is similar to the
construction of the \emph{restricted chase} of a database
described in~\cite{JoKl84}. \\
\textbf{Example 7:}Let us consider the following
Global-and-Local-As-View (GLAV)
   case when each dependency
   in $\M$ will be a \emph{tuple-generating dependency (tgd)} of the
   form\\
   $ ~(2)~~~~~~\forall {\bf x}~(\exists {\bf y}~q_{\S}(\bf x,\bf y)~ \Longrightarrow
   ~\exists {\bf z} ~q_{\G}(\bf x,\bf z))$\\
     where the formula $q_S(\bf x)$ is a conjunction of atomic
     formulas over $\S$ and $q_G(\bf x,\bf z)$ is a conjunction of
     atomic formulas over $\G$. Moreover, each target dependency
     in $\Sigma_T$ will be either a \emph{tuple-generating dependency (tgd)} of the form\\
     $~(3)~~~~~ \forall {\bf x} ~ (\exists {\bf y}~\phi_G(\bf x,\bf y)~ \Longrightarrow ~
      \exists {\bf z} ~ (\psi_G(\bf x,\bf z))$
     \\
     (we will consider only class of weakly-full tgd for which
     query answering is decidable, i.e., when the right-hand side has no existentially
      quantified variables, and if each $y_i \in \bf y$ appears at most once in the left
      side),\\
     or an \emph{equality-generating dependency (egd)}:\\
     $ ~(4)~~~~~\forall {\bf x} ~(\phi_G(\bf x)~ \Longrightarrow ~(x_1 = x_2 )) $\\
     where the formulae $\phi_G(x)$  and $\psi_G(x,y)$ are conjunctions of
     atomic formulae over $\G$, and $x_1, x_2~$ are among the
     variables in x.
     \\$\square$\\
     Notice that this example includes as special cases both LAV
     (when each assertion is of the form $q_S(\bf x)=~s(\bf x)$,
     for some relation $s$ in $\S$ and $\Map{q_{\S}}{q_{\G}}$)  and GAV
     (when each assertion is of the form $q_G(\bf x,\bf z)=~g(\bf x,\bf z)$,
     for some relation $g$ in $\G$ and $\Map{q_{\G}}{q_{\S}}$)
    data integration mapping in which the views are sound.
 \subsection{A categorial semantics of  database integrity constraints}
 It is natural for a database schema
$(\A,\Sigma_A )$, where $\A$ is a schema and $\Sigma_A$ are the
database integrity constraints, to take $\Sigma_A$ to be a
\emph{tuple-generating dependency (tgd)} and
\emph{equality-generating dependency} (egd). These two classes of
dependencies together comprise the \emph{embedded implication
dependencies} (EID) ~\cite{Fagi82} which seem to include essentially
all of the naturally-occuring constraints on relational databases.\\
Let $(\A,\Sigma_A )$ be a database schema expressed in a language
$\L_D$ over an alphabet $\A_D$, where  $\A$ is a schema and
$\Sigma_A = \Sigma_A^{tgd} \bigcup \Sigma_A^{egd}$ are the
database integrity constraints (set of EIDs).\\
We can represent it by a schema mapping $\Sigma_A:\A \longrightarrow
\A$, and its denotation in $DB$ can be given by an arrow, as
follows:
\begin{propo}\label{prop:IntConstr} If for a database schema $(\A,\Sigma_A )$ there
exists a model (instance-database) $A$ that satisfies all integrity
constraints $\Sigma_A = \Sigma_A^{tgd} \bigcup \Sigma_A^{egd}$, then
there exists an interpretation R-algebra $\alpha$ and its extension,
a functor $\alpha^*:Sch(\A,\Sigma_A )\longrightarrow DB$, where
$Sch(\A,\Sigma_A )$ is the category derived from the graph (arrow)
$\Sigma_A:\A \longrightarrow \A$ (composed by the single node $\A$,
the arrow $\Sigma_A$ and  the identity arrow $id_{\A}:\A
\longrightarrow \A$ equal to an empty set of integrity constraints;
composition of arrows in this category corresponds to the union
operator), such that:
\begin{itemize}
  \item $\alpha^*(\A) \triangleq A$, $~~$(set of relations
  $\alpha(R_i)$ for each predicate symbol $R_i$ in a schema $\A$)
  \item $\alpha^*(id_{\A}) \triangleq id_A:A\longrightarrow A$,
  $~~$ (identity arrow in $DB$ of the object $A$)
  \item $\alpha^*(\Sigma_A) \triangleq(f_{tgd} \bigcup f_{egd}):A \longrightarrow
  A$, where:\\Let $R_{1i}$ be
   the set of predicate letters used in a query
  $q_{A_i}(\textbf{x})$ where $\|q_{A_i}(\textbf{x})\|$ is its obtained view, and $q_i \in O(R_{1i},r'_i)$ be mapped into a
  view computation   $\alpha(q_i)$ with
  $\alpha(\partial_1(q_i))= \alpha(r'_i) = \|q_{A_i}(\textbf{x})\|$, then
\begin{enumerate}
  \item for each i-th tgd $q_{A_i}(\textbf{x})\Longrightarrow \exists
  \textbf{y}~q_{A2_i}(\textbf{x},\textbf{y})$ in $\Sigma_A^{tgd}$,
 we introduce a new predicate symbol $r_i$ with the interpretation
  $\alpha(r_i) = \|q_{A2_i}(\textbf{x},\textbf{y})\|$ (the view of $A$
  obtained from a query $q_{A2_i}(\textbf{x},\textbf{y})$ ), and\\
  $f_{tgd}\triangleq is^{-1}_A \circ \bigcup_{\partial_0(v_i) = \{r_i \}~\&~
~\partial_1(v_i) = \{r'_i\}}~\alpha(v_i \cdot q_i)~:A\longrightarrow A$\\
  where  $\alpha(v_i)$ is an inclusion-case tuple-mapping function in \ref{def:atomicmorphisms}.
%
  \item for each i-th egd $q_{A_i}(\bf x)~ \Longrightarrow ~(x_1 = x_2)$ in $\Sigma_A^{egd}$,
 we introduce a new predicate symbol $r_i$ with the interpretation $\alpha(r_i) = \|q_{A_i}(\textbf{x})\|$
 and\\
$f_{egd}\triangleq is^{-1}_A \circ \bigcup_{\partial_0(q_{Y_i}) =
\alpha(r_i)~\&~\partial_1(q_{Y_i}) = \{\perp\} } ~q_{Y_i}\circ
\alpha(q_i)~:A \longrightarrow A~$\\
where $~q_{Y_i}:TA \longrightarrow TA$ is a $Yes/No$ arrow in $DB$,
and $\alpha(q_i):A \rightarrow TA$ a view-map arrow in $DB$.
\end{enumerate}
  $is_A:A \simeq TA$ is an
isomorphism in $DB$ category, and $is^{-1}_A$ its inverse arrow.
\end{itemize}
\end{propo}
\textbf{Proof:} It is easy to verify that if $\alpha^*$ satisfies
the conditions in points 1 and 2, then all constraints in $\Sigma_A$
are satisfied, so that this functor is a Lavwere's model of a $\A$.
Notice that for a $Yes/No$ arrow in $DB$ category $~~q_{Y_i}:TA
\longrightarrow TA$, the $\partial_1(q_{Y_i}) = \perp^0 $ means that
for a view $\alpha(r_i) = \|q_{A_i}(\textbf{x})\|$ holds $(x_1 =
x_2)$, i.e., the answer of the query $q_{A_i}(\bf x)~
\Longrightarrow ~(x_1 = x_2)$ is $Yes$, and $\widetilde{f_{egd}} =
\bot^0$, for each egd constraint in $\Sigma_A^{egd}$.
\subsection{GLAV Categorial semantics}
Let us consider the most general case of GLAV mapping:
\begin{definition} \label{def:GLAV}  For a general GLAV data
integration/exchange system $\I = \tup{\B,\A,\M}$, when each tgd
 maps a view of one  database into a view of
 another database, we define the following two schema mappings,
$~f_A:\A\longrightarrow \C$, $~f_B:\B\longrightarrow \C$, where
$~\C$ is a new logical schema composed  by a new predicate symbol
$r_i(\bf x)~$ for a formulae $~q_{\B}(\bf x,\bf z)$, for every i-th
tgd $~\forall {\bf x}~(\exists {\bf y}~q_{\A}(\bf x,\bf y)~
\Longrightarrow   ~\exists {\bf z} ~q_{\B}(\bf x,\bf z)~$ in $\M$:
\[
  \ f_A ~\triangleq \bigcup_{ \partial_0(q_i)= R_{1i}
   ~\&~ \partial_1(q_i)= \partial_0(v_i)
   ~\&~ \partial_1(v_i) ~=~ \{ r_i\}} \{v_i\cdot q_i~: \A\longrightarrow
   \C\}\]
\[
  \ f_B ~\triangleq \bigcup_{ \partial_0(q_{B_i})= R_{2i}
   ~\&~ \partial_1(q_{B_i}) ~=~  \{r_i\}}  \{q_{B_i}~:\B\longrightarrow
   \C\}
\]
($R_{1i}$, $R_{2i}$  are, respectively, the set of predicate symbols
used in the query $q_{\A}(\bf x, \bf y)$ and the
  set of predicate letters used in the query $q_{\B}(\bf x,\bf z)$)
\end{definition}
 Note: in the particular cases (GAV and LAV), when a view of one
 database is mapped into one \emph{element} of another database,
 we obtain only a mapping arrow between two schemas.
 In fact in $\M:\A \longrightarrow \B$, for GAV
 a schema $\A$ is the source database and $\B$ is the global schema;
 for LAV it is the opposite.\\
We can generalize this framework into a complex  data
integration/exchange system $\I = \tup{\B_k,\A_k,\M_k}, k \in
N$.\\
Let $Sch(\I)$ be the category generated by the sketch (enriched
graph) $\I$. We can now define a mapping functor from the
scheme-level category into the instance level category $DB$:
\begin{theo}\label{th:GLAV}
If for each $~\tup{\B_k,\A_k,\M_k}$,  of the data
integration/exchange system $\I = \tup{\B_k,\A_k,\M_k}, k \in N$,
for a given instance A of the schema $\A$ there exists the universal
(canonical) instance $B = \can(\I,\D)$ of the global schema $\B$
legal w.r.t. A, then there exists  the interpretation R-algebra
$\alpha~$ and its extension, the functor (categorial Lawvere's
model) $\alpha^*:Sch(\I)\longrightarrow DB$, defined as
follows:\\
For every single data integration/exchange system $\tup{\B,\A,\M}$):
\begin{enumerate}
  \item for any schema arrow $f_B:\B \longrightarrow \C$ in
  $Sch(\I)$ it holds that
$ ~~~~B = \alpha^*(\B)\triangleq \can(\I,\D)$, and $ C =
\alpha^*(\C)$ is the database instance of the schema
  $\C$ composed by: for each i-th tgd $~\forall {\bf x}~(\exists {\bf y}~q_{\A}(\bf x,\bf y)~
\Longrightarrow   ~\exists {\bf z} ~q_{\B}(\bf x,\bf z)~$ in $\M$ we
have an element $\alpha(r_i) = \pi_X(\|q_{\B}(\bf x,\bf z)\|) $
   (the projection on $x$ of the view obtained from the query $q_{\B}(\bf x,\bf  z)~$ over $B = \can(\I,\D)$)
   in C, so that
\[
  \ \alpha^*(f_B) ~\triangleq \bigcup_{ \partial_0(q_{B_i})= R_{2i}
   ~\&~ \partial_1(q_{B_i}) ~=~  \{r_i\}}  \{\alpha(q_{B_i})~:B\longrightarrow
   C\}
\]
($R_{1i}$, $R_{2i}$  are, respectively, the set of predicate letters
used in the query $q_{\A}(\bf x, \bf y)$ and the
  set of predicate letters used in the query $q_{\B}(\bf x,\bf z)$);\\
  and for any schema arrow $f_A:\A \longrightarrow \C$ in
  $Sch(\I)$, it holds:
  $ A = \alpha^*(\A)$ is a given instance of the source schema
  $\A$, and
  \[
  \ \alpha^*(f_A) ~\triangleq \bigcup_{ \partial_0(q_i)= R_{1i}
   ~\&~ \partial_1(v_i) ~=~  \{r_i\}} \{\alpha(v_i \cdot q_i)~:A \longrightarrow
   C\}
\]
 where $ \alpha(v_i):\alpha( \partial_1(q_i))\longrightarrow \alpha(r_i)~$
 (with $ \alpha( \partial_1(q_i)) = \pi_X(\|q_{\A}(\bf x)\|)$  is the projection on $x$ of the
 view obtained from the query $q_{\A}(\bf x, \bf y)$~) is a function:
\begin{itemize}
  \item inclusion case, if i-th tgd has the same direction of its
  implication symbol (w.r.t arrow $f_A$)
  \item inverse-inclusion case, if i-th tgd has the opposite direction of its
  implication symbol
  \item equal case, if i-th tgd is an equivalence dependency
  relation.
\end{itemize}
\item Let  $~f^{inv}_A:C\longrightarrow A$ be the equivalent reverse
arrow of $\alpha^*(f_A) $ and $~f^{inv}_B:C\longrightarrow B$ be the
equivalent reverse arrow of $\alpha^*(f_B)$, then,  for each system
$\tup{\B,\A,\M}$) we obtain the equivalent direct mapping morphisms
$f = ~f^{inv}_B \circ \alpha^*(f_A):A\longrightarrow B~$ and $
f_{inv} = ~f^{inv}_A \circ \alpha^*(f_B) :B\longrightarrow A$ in DB
category.
\end{enumerate}
\end{theo}
\textbf{Proof:} Directly from the mapping properties of $DB$
morphisms and from the equivalent reversibility of its morphisms:
each morphism in $DB$ represents a denotational semantics for a well
defined exchange problem between two database instances, so we can
define a functor for such an exchange problem. Such a functor,
between the schema integration level (theory) and the instance level
(which is a model of this theory) is just an extended interpretation
function of a particular model of R-algebra.
\\$\square$\\
\textbf{Remark:}  A solution for a data integration/exchange system
does not exist always (if there exists a failing finite chase, see
~\cite{CCGL02,FKMP03} for more information), but if it exists then
it is a \emph{canonical universal solution} and in that case there
also exists  a mapping functor of the theorem above. So, this
theorem can be abbreviated by: " given a data exchange problem graph
$\I = \tup{\B_k,\A_k,\M_k},k \in N$, then:\\
$\exists \alpha^*:Sch(\I)\longrightarrow DB~~~~\textbf{iff}~~~~$
there exists a universal (canonical) solution for a correspondent
data integration/exchange problem".\\
The theorem above shows how GLAV mapping can be equivalently
represented by LAV and GAV mappings and shows that the query
answering under IC's can be done in the same way in LAV and GAV
systems.
\subsection{Query rewriting in GAV with (foreign) key constraints}
The characteristics of the components of a data integration system
in this approach ~\cite{CCGL02} are as follows:
\begin{itemize}
\item The \emph{global schema}, expanded by the \emph{new unary predicate} $Val(\_)$ such that
 $Val(c)$ is true if $c \in \textbf{dom}$, is expressed in the relational model with  $\Sigma_T$
 (key and foreign key constraints). We assume that in such a global schema $\G$ there is
  exactly one key constraint for each relation.
\begin{enumerate}
  \item \emph{Key constraints}: given a relation $r$ in the schema, a key
  constraint over $r$ is expressed in the form $\mathit{key}(r) = \ins{At}$,
  where $\ins{At}$ is a set of attributes of $r$. Such a constraint is satisfied
  in an instance-database $A$ if for each $t_1,t_2\in\ext{r}{A}$, with $t_1\neq t_2$,
  we have $t_1[\ins{At}]\neq t_2[\ins{At}]$, where $t[\ins{At}]$ is the projection
  of the tuple $t$ over $\ins{At}$.
\item \emph{Foreign key constraints}: a foreign key constraint is a statement
  of the form $r_1[\ins{At}] \subseteq r_2[\ins{Bt}]$, where $r_1,r_2$ are
  relations, $\ins{At}$ is a sequence of distinct attributes of $r_1$, and
  $\ins{Bt}$ is $\key(r_2)$, i.e., the sequence $[1,\ldots,h]$
  constituting the key of $r_2$.  Such a constraint is satisfied in a database
  $A$ if for each tuple $t_1$ in $\ext{r_1}{A}$ there exists a tuple
  $t_2$ in $\ext{r_2}{A}$ such that $t_1[\ins{At}]=t_2[\ins{Bt}]$.
\end{enumerate}
\item The \emph{mapping} $\M$ is defined following the GAV (global-as-view) approach:
  to each relation $r$ of the global schema $\G$ we associate a query $\rho(r)$
  over the source schema $\S$: we assume that this query preserves the key constraint of $r$.
  \item For each relation $r$ of the global schema, we may compute the relation
  $r^\D$ by evaluating the query $\rho(r)$ over the source database $\D$, and  compute the
  relation $Val$ for all constants in \textbf{dom}.
  The various relations so obtained define what we call the \emph{retrieved global
   database} $~\ret(\I,\D)$.  Notice that, since we assume that $\rho(r)$ has been
  designed so as to resolve all key conflicts regarding $r$, the retrieved
  global database satisfies all key constraints in $\G$.
\end{itemize}
\begin{figure}
\begin{center}
 \includegraphics{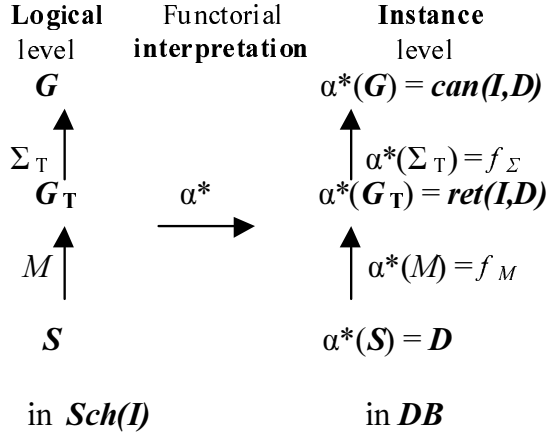}
 \caption{Functorial translation}
 \label{fig-functorial-translation}
 \end{center}
 \end{figure}
In our case, \emph{with integrity constraints} and with \emph{sound
mapping}, the semantics of a data integration system $\I$  is
specified in terms of a set of \emph{legal} global
instance-databases, namely, those databases (they exits iff $\I$
\emph{is consistent} w.r.t. $\D$, i.e., iff $\ret(\I,\D)$ does not
violate any key constraint in $\G$) that are supersets of the
\emph{retrieved} global database $\ret(\I,\D)$. \\
  In ~\cite{CCGL02}, given the retrieved global database $\ret(\I,\D)$, we may construct
inductively the canonical database $\can(\I,\D)$  by starting from
$\ret(\I,\D)$ and repeatedly applying the following rule:
\begin{quote}
  if $(x_1,\ldots,x_h)\in r^{\can(\I,\D)}[\ins{A}]$, $(x_1,\ldots,x_h)\not\in
  r_2^{\can(\I,\D)}[\ins{B}]$, and the foreign key constraint
  $r_1[\ins{A}] \subseteq r_2 [\ins{B}]$ is in  $\Sigma_T \subseteq \G$,\\
  then insert in $r_2^{\can(\I,\D)}$ the tuple $t$ such that
  \begin{itemize}
  \item $t[\ins{B}] = (x_1,\ldots,x_h)$, and
  \item for each $i$ such that $1\leq i\leq\arity{r_2}$, and $i$ not in
    $\ins{B}$, $t[i]=f_{r_2,i}(x_1,\ldots,x_h)$.
  \end{itemize}
\end{quote}
 Notice that  the above rule  does enforce
the satisfaction of the foreign key constraint $r_1[\ins{A}]
\subseteq r_2 [\ins{B}]$ by adding a suitable tuple in $r_2$: the
key of the new tuple is determined by the values in $r_1[\ins{A}]$,
and the values of the non-key attributes are formed by means of the
Skolem function symbols $f_{r_2,i}$.\\
Based on the results in ~\cite{CCGL02}, $\can(\I,\D)$ is an
appropriate database for answering queries in a data integration
system. Notice that the terms involving Skolem functions are never
part of \emph{certain answers}. Thus, the lifted queries $q$ use the
$Val(\_)$ predicate
in order to eliminate the tuples with a Skolem values in $\can(\I,\D)$. \\
 Consequently, at the logic level, this GAV data
integration system can be represented by the graph composed by two
arrows (Figure \ref{fig-functorial-translation}) , $\M:\S
\longrightarrow \G_T$ and $\Sigma_T:\G_T \longrightarrow
\G~~$($Sch(\I)$ denotes the category derived by this graph).
\begin{definition} \label{def:Ftranslation} \textsl{Functorial interpretation} of this logic
scheme into denotational semantic domain $DB$,
$\alpha^*:Sch(\I)\longrightarrow DB$,  is defined  by two
corresponding arrows (Fig.~\ref{fig-functorial-translation})\\
$f_M:D \longrightarrow \ret(\I,\D)$,
$f_{\Sigma}:\ret(\I,\D)\longrightarrow \can(\I,\D)$, where
$\alpha^*(\S) = \D$ is the extension of the source database $\D$,
$\alpha^*(\G_T)= \ret(\I,\D)$ is the retrieved global database,
$\alpha^*(\G)=\alpha^*(\G_T,\Sigma_T)=\can(\I,\D)$ is the universal
(canonical) instance of the global schema with the
integrity constraints, and\\
$f_M \triangleq \bigcup\{
~q_{D_i}~|~where~\partial_1(q_{D_i})\triangleq \{\rho^D(r)\}$,
$~\partial_0(q_{D_i})$ is the set of all predicate symbols in the
query $\rho(r)$,
 $~( \Map{\rho(r)}{ r})\in \M \}$\\
$f_{\Sigma} \triangleq \bigcup\{\alpha(v_k \cdot q_{ret_k})~|~$
$~\partial_0(q_{ret_k})=
\partial_1(q_{ret_k})=\{r'\} ~, ~r' \in \ret(\I,\D)$,
where  $\alpha(v_k)$ is an inclusion-case tuple-mapping function (in
\ref{def:atomicmorphisms}) for $r'\}$,\\ because $\ret(\I,\D)$ and
$\can(\I,\D)$ have the same set of predicate symbols, but the
extension of each of them in $\ret(\I,\D)$ is a subset of the
extension in $\can(\I,\D)$.
\end{definition}
\textbf{Query rewriting coalgebra semantics:}\\
  The naive computation  is impractical, because
it requires the building of a canonical database, which is generally
infinite. In order to overcome this problem, a \emph{query rewriting
algorithm} ~\cite{CCGL02} consists of two separate phases.
\begin{enumerate}
\item Instead of referring explicitly to the canonical database for  query
  answering, this algorithm transforms the original lifted query $q$ into a new query $\expand{q}$
  over a global schema, called the \emph{expansion of $q$ w.r.t. $\G$}, such
  that the answer to $\expand{q}$ over the retrieved global database is equal
  to the answer to $q$ over the canonical database.
\item In order to avoid the building of the retrieved global database, the query does not
  evaluate $\expand{q}$ over the retrieved global database.  Instead, this algorithm
  unfolds  $\expand{q}$ to a new query, called $\unfold{\expand{q}}$, over the source
  relations on the basis of $\M$, and then uses the unfolded query
  $\unfold{\expand{q}}$ to access the sources.
\end{enumerate}
\begin{figure}[tb]
  \centering
  \resizebox{.65\columnwidth}{!}{\input{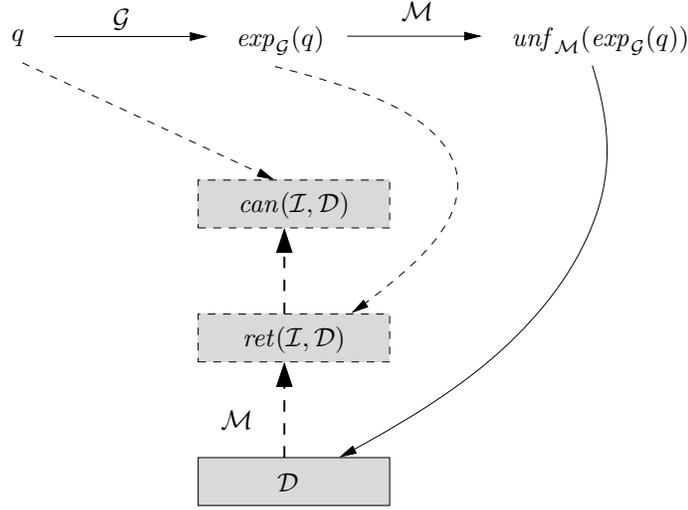}}
  \caption{Query answering process}
  \label{fig-query-processing}
\end{figure}
Figure ~\ref{fig-query-processing} shows the basic idea of this
approach (taken from ~\cite{CCGL02}). In order to obtain the
\emph{certain answers} $q^{\I,\D}$, the user lifted query $q$ could
in principle be evaluated (dashed arrow) over the (possibly
infinite) canonical database $\can(\I,\D)$, which is generated from
the retrieved global database $\ret(\I,\D)$.  In turn, $\ret(\I,\D)$
can be obtained from the source database $\D$ by evaluating the
queries of the mapping. This query answering process instead expands
the query according to the constraints in $\G$, than unfolds it
according to $\M$, and then evaluates it on the
source database.\\
Let us show how the symbolic diagram in
Fig.~\ref{fig-query-processing} can be effectively represented by
commutative diagrams  in $DB$, correspondent to the homomorphisms
between T-coalgebras representing equivalent queries over these
three instance-databases:  each query in $DB$ category is
represented by an arrow, and  \emph{can be composed} with arrows
that semantically denote mappings and integrity constraints.
\begin{theo}
Let $\I=\tup{\G,\S,\M}$ be a data integration system , $\D$  a
source  database for $\I$, $\ret(\I,\D)$ the retrieved global
database for $\I$ w.r.t. $\D$ , and $\can(\I,\D)$ the universal
(canonical) database for $\I$ w.r.t. $\D$.\\
Then, a denotational semantics for query rewriting algorithms
$\expand{q}$  and $\unfold{q}$, for a query expansion and query
unfolding respectively, are given by two (partial) functions on
T-coalgebras:
\begin{center}
$\unfold{\_} \triangleq Tf_M^{inv}\circ \_ \circ f_M $\\
$\expand{\_} \triangleq Tf_{\Sigma}^{inv}\circ \_ \circ
f_{\Sigma}$ and\\
$\unfold{\expand{\_}}\triangleq T(f_{\Sigma}\circ f_M)^{inv}\circ \_
\circ (f_{\Sigma}\circ f_M) $
\end{center}
where $f_M $ and $f_{\Sigma}$ are given by a functorial translation
of the mapping $\M$ and integrity constraints $\Sigma_T$.
\end{theo}
\textbf{Proof:} Let us denote by $q_E = \expand{q}$ and $q_U =
\unfold{\expand{q}}$ the  expanded and successively unfolded queries
of the original lifted query $q$. Then, by the query-rewriting
theorem  the diagrams
\begin{diagram}
 TD        & \rTo^{Tf} & T\ret(\I,\D) & \rTo^{Tf_{\Sigma}} & T\can(\I,\D)\\
 \uTo^{q_U}&           & \uTo^{q_E}    &                   &   \uTo_q \\
 D         & \rTo^{f}  &  \ret(\I,\D) & \rTo^{f_{\Sigma}}  &  \can(\I,\D)\\
\end{diagram}
 based on the composition of
T-coalgebra homomorphisms $f_M:(\D,q_U)\longrightarrow
(\ret(\I,\D),q_E)$ and $f_{\Sigma}:(\ret(\I,\D),q_E) \longrightarrow
\can(\I,\D)$, commute. It is easy to verify the first two facts.
Then, from the composition of these two
functions, we obtain\\
$\unfold{\expand{\_}} = \unfold{\_}\expand{\_} =Tf_M^{inv}\circ
(\expand{\_}) \circ f_M = Tf_M^{inv}\circ (Tf_{\Sigma}^{inv}\circ \_
\circ f_{\Sigma}) \circ f_M = (Tf_M^{inv}\circ
Tf_{\Sigma}^{inv})\circ \_ \circ (f_{\Sigma}) \circ f_M) =
T(f_{\Sigma}\circ f_M)^{inv}\circ \_ \circ (f_{\Sigma}\circ f_M)$
because of the duality and functorial property of $T$.
\\$\square$
%
\subsection{Fixpoint operator for finite canonical solution}
The database instance $\can(\I,\D)$ can be an infinite one (see an
example bellow), thus impossible to materialize for real
applications. Thus, in this paragraph we introduce a new approach to
the canonical model,  closer to the data exchange approach
~\cite{FKMP03}. It is not restricted to the existence of
query-rewriting algorithms, and thus can be used in order to define
a Coherent Closed World Assumption for data integration systems also
in the absence of query-rewriting algorithms \cite{Majkw04}. The
construction of the canonical\emph{ model} for a global schema of
the logical theory $\P_{\G}$ for a data integration system is
similar to the construction of the canonical \emph{database}
$\can(\I,\D)$ described in ~\cite{CCGL02}. The \emph{difference}
lies in the fact that, in the construction of this revisited
canonical model, denoted by $\can_M(\I,\D)$, for a global schema,
fresh \emph{marked null values} (set $SK = \{ \omega_0,
\omega_1,...\}$ of Skolem constants) are used instead of terms
involving Skolem functions,
 following the idea of construction of the restricted chase of a database described in ~\cite{JoKl84}.
 Thus, we enlarge a set of ordinary constants  \textbf{dom} of our language by $\Gamma_U = \textbf{dom} \bigcup SK$. \\
Another motivation for concentrating on canonical models is a view
~\cite{Reit78} that many logic programs are appropriately thought of
as having two components, an \emph{intensional} database (IDB) that
represents the reasoning component, and the \emph{extensional}
database (EDB) that represents a collection of facts. Over the
course of time, we can "apply" the same IDB to many quite different
EDBs. In this context it make sense to think of the IDB as
implicitly defining a transformation from an EDB to a set of derived
facts: we would like the set of derived
facts to be the canonical model.\\
 Now we construct inductively the revisited canonical database
model $\can_M(\I,\D)$ over the domain $\Gamma_U$ by starting from
$\ret(\I,\D)$ and repeatedly applying the following rule:
\begin{quote}
  if $(x_1,\ldots,x_h)\in r_1^{\can_M(\I,\D)}[\ins{A}]$, $(x_1,\ldots,x_h)\not\in
  r_2^{\can_M(\I,\D)}[\ins{B}]$, and the foreign key constraint
  $r_1[\ins{A}] \subseteq r_2 [\ins{B}]$ is in  $\G$,\\
  then insert in $r_2^{\can_M(\I,\D)}$ the tuple $t$ such that
  \begin{itemize}
  \item $t[\ins{B}] = (x_1,\ldots,x_h)$, and
  \item for each $i$ such that $1\leq i\leq\arity{r_2}$, and $i$ not in
    $\ins{B}$, $t[i]= \omega_k$, where $\omega_k$ is a fresh marked null value.
  \end{itemize}
\end{quote}
Note that the above rule does enforce the satisfaction of the
foreign key constraint $r_1[\ins{A}] \subseteq r_2 [\ins{B}]$, by
adding a suitable tuple in $r_2$: the key of the new tuple is
determined by the values in $r_1[\ins{A}]$, and the values of the
non-key attributes are formed by means of the fresh marked values
$\omega_k$ during the
application of the rule above.\\
The rule above defines the "immediate consequence" monotonic
operator $T_B$ defined by:\\
$~~~~~~T_B(I) = I \bigcup~\{~A ~|~~~A \in B_{\G}  , A \leftarrow
A_1\wedge
..\wedge A_n $ is a ground instance of \\
 $~~~~~~~~~~~~~~~~~~~~~~~~~~~~~~~~~~~~~~~~~~~~~~~~$ a rule in $\Sigma_{\G}$ and $\{A_1,..,A_n\}\in I~~\}$\\
where, at the beginning $I = \ret(\I,\D)$, and $B_{\G}$ is a
Herbrand base for a global schema. Thus, $\can_M(\I,\D)$ is a least
fixpoint
of this immediate consequence operator.\\
\textbf{Example 8:}  Suppose that we have two relations $r$ and $s$
in $\G$, both of arity $2$ and
  having as key the first attribute, and that the following dependencies hold
  on $\G$: \\ $~~ ~~ r[2] \subseteq s[1],~~ ~~s[1] \subseteq
  r[1]$.\\
  Suppose that the retrieved global database stores a single tuple $(a,b)$ in
  $r$.  Then, by applying the above rule, we insert the tuple $(b,\omega_1)$
  in $s$; successively we add $(b,\omega_2)$ in $r$, then
  $(\omega_2, \omega_3)$ in $s$, and so on.  Observe that the two
  dependencies are cyclic, and in this case the construction of the canonical
  database requires an infinite sequence of applications of the
  rules.
     The following table represents the computation of canonical database:
\begin{center}
\begin{tabular}{|c|c|}
  \hline
  $$ & $$ \\
   $~~r^{\can_M(\I,\D)}$ & $~~s^{\can_M(\I,\D)}$ \\
  \hline
  $$ & $$ \\
   $a,b$ & $b,\omega_1$ \\
   $b,\omega_2$ & $\omega_2,\omega_3$ \\
   $\omega_2,\omega_4$ & $\omega_4,\omega_5$ \\
   $\omega_4,\omega_6$ & $\omega_6,\omega_7$ \\
   $..$ & $..$ \\ \hline
\end{tabular}
\end{center}
 Thus, the canonical model $\can_M(\I,\D)$ is a legal database
model for the global schema.\\
Each \emph{certain answer} of the original user query
$q(\textbf{x})$, $\textbf{x} = \{x_1,..,x_k \}$ over a global schema
is equal to the answer $q_L(\textbf{x})^{\can_M(\I,\D)} $  of the
\emph{lifted} query $q_L(\textbf{x}) \equiv q(\textbf{x}) \wedge
Val(x_1) \wedge... \wedge Val(x_k)$ over this canonical model. Thus,
if  it were possible to materialize this canonical model, the
certain answers could be obtained over such a database. Often it is
not possible because (as in the example above) this canonical model
is \emph{infinite}. In that case, we can use the revisited fixpoint
semantics described in ~\cite{Majk03f}, based on the fact that,
after some point, the new tuples added into a canonical model insert
only new Skolem constants which are not useful  in order to obtain
\emph{certain} answers (true in all models of a database). In fact,
Skolem constants are not part of any certain answer to conjunctive
query. Consequently, we are able to obtain a \emph{finite subset} of
a canonical \emph{database},
which is large enough to obtain all certain answers. \\
Let us denote such a finite database by $\C_M(\I,\D)$, where\\ $r =
\{(a,b),(b,\omega_2), (\omega_2,\omega_4)\}$, $s = \{(b,\omega_1),
(\omega_2,\omega_3) \}~~~$ is a finite least fixpoint  which can be
used in order to obtain certain
answers to lifted queries.\\
  $\square$\\
In fact, we introduced marked null values (instead of Skolem
functions) in order to define and materialize such a \emph{finite}
database: it \emph{is not} a model of the data integration system
(which is infinite), but has all necessary query-answering
properties: it is able to give all certain answers to conjunctive
queries over a global schema.
Thus it can be materialized and used for query answering, instead of query-rewriting algorithms.\\
%
 The procedure for computation of a canonical database for the global
 schema, based on "immediate consequence" monotonic
operator $T_B$ defined in precedence, can be intuitively described
as follows: it starts with an instance $<I,\emptyset>$ which
consists of $I$, instance of the source schema, and of the empty
instance $\emptyset$ for the target (global schema). Then we chase
$<I,\emptyset>$ by applying all the dependencies in $\Sigma_{st}$ (a
finite set of source-to-target dependencies) and $\Sigma_{t}$ (a
finite set of target integrity dependencies)  as long as they are
applicable. This process may fail (if an attempt to identify two
domain constants is made in order to define a homomorphism between
two consecutive target instances) or it may never terminate. Let
$J_i$ and $J_{i+1}$ denote two consecutive target instances of this
process ($J_0 = \emptyset$), then we introduce a function
$C_h:\Theta \longrightarrow \Theta$, where $\Theta$ is the set of
all pairs $<I,J>$, $I$ is a source instance and $J$ one
generated by $I$ target instances, such that:\\
$ <I,J_{i+1}> = C_h(<I,J_i>) \supseteq <I,J_i>$\\
This function is  monotonic.
Let us define the sets\\
$S_i = T_w(\pi_2(<I,J_i>)) = T_w(J_i)$\\
and the fixpoint operator $\Psi:\Theta_w\longrightarrow \Theta_w$,
where $\Theta_w = \{~T_w(\pi_2(S))~|~S \in \Theta \}$, such that
$\Psi(T_w(\pi_2(<I,J_i>))) = T_w(\pi_2(C_h(<I,J_i>)))$, i.e.,$\Psi
T_w\pi_2 = T_w \pi_2 C_h:\Theta\longrightarrow \Theta_w$, and with
the least fixpoint $\C_M(\I,\D)= S$,  $ S = \Psi(S)$.\\
\begin{propo} \cite{Majk03f} Let $<I,\emptyset>$ be an initial instance
that consists of $I$, a finite instance of the source schema, and of
an empty instance $\emptyset$ for the target (global schema). Then,
there exists the least fixpoint $S$ of the function
$\Psi:\Theta_w\longrightarrow \Theta_w$, which is equal to $S = T_w
\pi_2C_h^n(<I,\emptyset>)$ for a finite $n$.
 \end{propo}
Consequently, we can demonstrate the following algebraic property
for the closure operator $T_w$:
\begin{propo} The closure operator $T_w$ is algebraic, that is,
given any infinite canonical database $\can(\I,\D)$, holds that
\[ T_w( \can(\I,\D)) = \bigcup \{ T_w(X')~|~X' \subseteq_\omega
\can(\I,\D)\}\] where $X' \subseteq_\omega \can(\I,\D)\}$ means that
$X'$ is a \textsl{finite} subset of $\can(\I,\D)$.
\end{propo}
\textbf{Proof:} In fact, for $X' =\pi_2C_h^n(<I,\emptyset>)$ for a
finite $n$ and, consequently, finite $X'$, such that $X'$ is the
least fixpoint of $\Psi$, i.e., $ X' = \Psi(X')$, holds that $T_w(
\can(\I,\D)) = T_w(X')$.
\\$\square$\\
Notice that each infinite canonical database of a global database
schema $\G$ is weakly equivalent to
 its finite subset (an instance-database) $\C_M(\I,\D) = X'$, where $ X' =
 \Psi(X')$ is a finite subset of $\can(\I,\D)$, that \emph{is not a model} of $\G$  but  is obtained as the least fixpoint of the
 operator $ \Psi$.\\ Thus, $~~\can(\I,\D)~\approx_w~\C_M(\I,\D)$,
 where $~\can(\I,\D)$ is an infinite model of $\G$, and
 $~\C_M(\I,\D)$ is a finite weakly equivalent object
 to it in $DB$ category.

\section{Conclusion}
We have presented only a fundamental overview of a new approach to
the database concepts developed from an observational equivalence
based on views. The main intuitive result of obtained basic database
category $DB$, more appropriate than the category $Set$ used for
categorial Lawvere's theories, is to have the possibility of making
synthetic representations of database mappings, and queries over
databases in a graphical form, such that all mapping (and query)
arrows can be composed in order to obtain the complex database
mapping diagrams. Let us consider, for example, the P2P systems or
mappings between databases in a complex Datawarehouse. Formally, it
is possible to develop a graphic (sketch-based) tool for a
meta-mapping description of complex (and partial) mappings in
various contexts, with a formal
mathematical background.\\
 These, and some other, results suggest the need for further investigation of:
\begin{itemize}
  \item The semantics for Merging and Matching database operators based
  on a complete database lattice, as in ~\cite{BuDK92}.
  \item The expressive power of the $DB$ category with Universal
  Algebra considerations.
  \item Monad based consideration of category $DB$ as a computation
  model for view-based database mappings.
  \item A complete investigation of all paradigms for
  database mappings .
  \item A formalization in this context of query processing in a P2P framework
\end{itemize}
We still have not  considered other important properties of this
$DB$ category, such as algebraic properties for finitary
representation of infinite databases, that is, locally finitely
representation properties \cite{AdRo94}, or monoidal enrichments,
based on concept of matching of two databases, which can be used for
enriched Lawvere-s theories of sketches \cite{KiPT99,GoPo98,Powe00}
in very-expressive database algebraic specification for complex
inter-database mappings.\\

\textbf{Acknowledgments:} I warmly thank Eugenio Moggi and Giuseppe
Rossolini for their invitation to a seminar at DISI Computer
Science, University of Genova, Italy, December 2003, and for a
useful discussion and correspondence that have given me the
opportunity to improve an earlier version of this  work.


\bibliographystyle{IEEEbib}
\bibliography{mydb}


\end{document}